\documentclass[nofootinbib]{revtex4-1}
\usepackage{amsmath}    
\usepackage{graphicx}   
\usepackage{verbatim}   
\usepackage{color}      
\usepackage{subfigure}  
\usepackage{hyperref}   
\usepackage{multirow}

\begin{document}

\begin{abstract}
We present a new stochastic approach to describe and remodel the conversion process of a wind farm at a sampling frequency of 1Hz. The method is trained on data measured on one onshore wind farm for an equivalent time period of 55 days. Three global variables are defined for the wind farm: the 1-Hz wind speed $u(t)$ and ten-minute average direction $\bar{\phi}$ both averaged over all wind turbines, as well as the cumulative 1-Hz power output $P(t)$. When conditioning on various wind direction sectors, the dynamics of the conversion process $u(t) \to P(t)$ appear as a fluctuating trajectory around an average IEC-like power curve, see section \ref{sec:wf}. Our approach is to consider the wind farm as a dynamical system that can be described as a stochastic drift/diffusion model, where a drift coefficient describes the attraction towards the power curve and a diffusion coefficient quantifies additional turbulent fluctuations. These stochastic coefficients are inserted into a Langevin equation that, once properly adapted to our particular system, models a synthetic signal of power output for any given wind speed/direction signals, see section \ref{sec:model}. When combined with a pre-model for turbulent wind fluctuations, the stochastic approach models the power output of the wind farm at a sampling frequency of 1Hz using only ten-minute average values of wind speed and directions. The stochastic signals generated are compared to the measured signal, and show a good statistical agreement, including a proper reproduction of the intermittent, gusty features measured. In parallel, a second application for performance monitoring is introduced in section \ref{sec:monitor}. The drift coefficient can be used as a sensitive measure of the global wind farm performance. When monitoring the wind farm as a whole, the drift coefficient registers some significant deviation from normal operation if one of twelve wind turbines is shut down during less than $4\%$ of the time. Also, intermittent anomalies can be detected more rapidly than when using ten-minute averaging methods. Finally, a probabilistic description of the conversion process is proposed and modeled in appendix \ref{sec:proba}, that can in turn be used to further improve the estimation of the stochastic coefficients.

\vspace*{2ex}\noindent
Published as:\\
P.~Milan, M.~W{\"a}chter, J.~Peinke, Stochastic modeling and
performance monitoring of wind farm power production, J.~Renewable
Sustainable Energy \textbf{6}, 033119 (2014)

\end{abstract}

\title{Stochastic modeling and performance monitoring of wind farm power production}
\author{Patrick Milan}
\author{Matthias W{\"a}chter}
\author{Joachim Peinke}
\affiliation{ForWind - Center for Wind Energy Research, Institute of Physics,\\University of Oldenburg, 26129 Oldenburg, Germany}
\maketitle

\section{Introduction}
\label{sec:intro}

Wind energy is currently the fastest growing energy sector in Europe \cite{WWEA2010}. The fast integration of the complex wind resource in electric networks raises new challenges in terms of grid stability \cite{Liu2011,Ayodele2012,Milan2013a}. Smart grid concepts are being developed to cope with the turbulent nature of wind power, that involve e.g. energy storage or smart control strategies. The challenge is on the one hand technical, as the current grid structure must evolve from a strongly polarized source/load configuration with few large power sources towards a delocalized distribution of small renewable sources \cite{Rohden2012}. New grid concepts must be designed to handle the increasing amount of fluctuating wind power, see e.g. Ref. \cite{Liu2013} for an example on active wind power regulation. On the other hand, a fundamental understanding of the wind resource still lacks. Wind energy planing still focuses mostly on large-scale meteorological wind dynamics. Dynamics at faster time scales in minutes are typically simplified to a Gaussian wind field \cite{IEC}, bypassing the complex dynamics of turbulence \cite{Lovejoy2001,Boettcher2007a,Morales2010a}. Yet it is observed on measured data that fast wind fluctuations deviate from Gaussianity \cite{Morales2010a}, and are instead intermittent multifractals \cite{Lovejoy2001,Lovejoy2009}. This observation must be coupled with the fact that modern wind turbine designs optimize power performance \cite{Bianchi2006} by following these fast, intermittent wind fluctuations. Besides the obvious impact on the mechanical loads acting on the turbine machinery \cite{Tavner2011,Muecke2011}, such performance-oriented control strategy implies feeding intermittent gusts into the power grid \cite{Milan2013a}. Our increasing dependence on wind energy stresses the necessity to reliably predict first the wind dynamics, and second the resulting wind power production.

The first aspect is a fundamental challenge that traditionally involves both fields of meteorology for large-scale dynamics and turbulence for fast fluctuations. The central problem is that wind dynamics involve a wide range of spatio-temporal scales that are entangled in an energy cascade. The standard picture involves a three-dimensional downward cascade at small scales and a two-dimensional inverse cascade at mesoscales. In between, a hypothetical spectral gap \cite{vdHoven1957} would separate the two regimes. Some recent studies contradict this hypothesis and propose a universal cascade model instead \cite{Schertzer2012,Fitton2011}. Besides solving directly Navier-Stokes equation including all external influences such as thermal transfers or topography (that is far from achievable), no simpler alternative is recognized as a valid method to describe the entire wind dynamics. Instead various meteorological models parametrize all the influencing effects, but they are limited in their resolution due to their high computational cost, nowadays resolving spatial and temporal scales of typically $1$ km$^2$ and $1$ hour. The smaller structures (that remain quite large) are usually not modeled using CFD, but with a statistical model of turbulence, e.g. a Gaussian correlated field \cite{IEC}. It should be noted that high-frequency wind models have been developed for decades \cite{Kaminsky1991}, and the turbulence community has proposed several approaches in recent years \cite{Muzy2010b,Calif2012a,Calif2012b}.

The second aspect consists in modeling the conversion process operated
by the wind installation, seen here as the conversion of a wind speed
$u$ conditioned on the wind direction $\phi$ into an electrical power
output $P$. The complete conversion process is commonly simplified to
an average power curve \cite{IEC-12} in the case of a single wind
turbine. Being an average curve, this approach is a good description
of the long-term behavior, but naturally fails to describe the
dynamics in the faster time scales of minutes \cite{Boettcher2007a}
\footnote{More accurate approaches involve an aero-mechanical
  description of each wind turbine \cite{Burton2001}, which becomes
  excessively demanding for everyday prediction.}. Ref. \cite{Wu2014}
recently proposed a prediction model for the low-frequency (30-minute)
trend of a wind farm power output from weather data using a mixed
statistical/CFD approach. It has been recently shown that the
cumulative power output of a wind farm \cite{Milan2013a} or a 300-km
large wind installation \cite{Kamps2012} contain intermittent
fluctuations at time scales of minutes and even seconds that are
typical of a multifractal process \footnote{This observation
  contradicts the intuitive yet false argument claiming that the
  fluctuations of neighboring wind turbines cancel out. Considering
  many wind installations separated by a distance shorter than the
  correlation length of atmospheric winds (in hundreds of kilometers
  according to Ref. \cite{Muzy2010a}), these {\it neighboring}
  installations are driven by correlated wind fields and produce
  correlated outputs. They are not independent power sources, and
  their cumulative output does not sum up to a Gaussian process, that
  is the central limit theorem cannot be applied.}. The standard power
curve method overlooks such effects, so it is impossible to predict
precisely how stable the power grid will be at the short time scales
at which it is operated.

For the optimal exploitation of wind energy, it is also important to minimize downtimes. Servicing and spare parts were found to account for one fourth of operation and maintenance costs for onshore installations, i.e. about $1\%$ of the total investment \cite{Blanco2009}. Ref. \cite{Hahn2007} made a 15-year-long study and showed that the 1500 onshore wind turbines studied had an average availability of $98\%$ (downtimes occurred during $2\%$ of the time). About half of the recorded failures were attributed to electrical and control systems, the other half coming from mechanical systems. It is interesting to note that failure rates were especially high for wind turbines of 1MW and more. Recent results \cite{IWES2012} show that the availability of offshore installations is much lower, in the order of $70-90\%$. This makes the development of procedures for early warning, identification and reparation of damages essential, see Ref. \cite{Ciang2008} for an overview of various detection methods. In particular, methods for the early prediction of emerging failures are very valuable in order to react optimally. There are strong indications that the turbulent nature of wind flows affects the mechanical fatigue of wind turbines \cite{Muecke2011}. This underlines the necessity to include the fast dynamics of turbulence in wind energy methods in order to further improve the reliability of wind energy systems.

In this paper, we propose an alternative approach that describes and remodels the conversion process of a wind farm at a frequency of $1$ Hz. Our analysis is performed on a wind farm that is described in section \ref{sec:wf}. We present a stochastic model for the conversion process in section \ref{sec:model}, where the dynamics are intuitively characterized through a set of stochastic estimates. We show in section \ref{sec:monitor} that these estimates are highly reactive to dynamical changes, and we promote them as well for condition monitoring of the power performance for a wind farm. Also, a probabilistic description of the power production is presented in appendix \ref{sec:proba}.

\section{Wind farm dynamics}
\label{sec:wf}

\subsection{Data description}
\label{sec:data}

All results presented in this paper were extracted from measurement data for one wind farm. The wind farm is installed onshore over an area covering roughly $4km^2$, and is surrounded by flat rural terrain. It consists of 12 identical variable-speed, pitch-regulated wind turbines. The rated power of each turbine is in the order of $2$MW. Its exact value cannot be published following an agreement with the farm manager. For this reason, all values of power output will be normalized to the rated power of the entire wind farm $P_r = 100\% \simeq 12 \times 2$MW.

Measurements were conducted synchronously at each wind turbine. The measured signals are the net electrical power output generated by each wind turbine, and the wind speed and direction measured on each nacelle by a cup anemometer and a wind vane. The operational status of each wind turbine was also provided, so that the data was rejected when one or more turbines were not operating in normal conditions.

All measurements were performed at a sampling frequency $f_s=1$Hz. Unless stated otherwise, all the time series presented in this paper have a sampling frequency of $1$Hz. The measurement campaign was conducted over a period of eight months, from June 2009 till February 2010. The measurements were regularly interrupted, eventually leaving $7,775$ ten-minute periods ($4,665,000$ samples, or $53$ days $23$ hours and $50$ minutes).

\subsection{Wind farm observables}
\label{sec:observables}

This paper aims towards wind farm dynamics, and not dynamics of single wind turbines. In order to describe the entire wind farm, some new observables were defined from the data measured on all single wind turbines. Ref. \cite{Cutler2011} observes that the average of all single wind speed and direction measurements is a more precise measure for power curve modeling than an eventual met mast measurement. Following this observation, we define three observables for the wind farm. The average wind speed over the entire wind farm is defined as
\begin{eqnarray}
u(t) =\frac{1}{N} \sum_{i=1}^N u_i(t) \, ,
\label{eq:u}
\end{eqnarray}
where $u_i(t)$ is the wind speed measured at $1$Hz by the cup anemometer on the nacelle of turbine $i$. Similarly, the average wind direction over the entire wind farm is defined following
\begin{eqnarray}
\phi(t) =\frac{1}{N} \sum_{i=1}^N \phi_i(t) \, ,
\label{eq:phi}
\end{eqnarray}
where $\phi_i(t)$ is the wind direction measured at $1$Hz by the wind vane on the nacelle of turbine $i$. Finally, the total power output fed by the wind farm into the electric grid is
\begin{eqnarray}
P(t) = \sum_{i=1}^N P_i(t) \, ,
\label{eq:P}
\end{eqnarray}
where $P_i(t)$ is the net electrical power output of turbine $i$ at $1$Hz. $N$ represents the total number of turbine measurements considered, in our case $N=12$ corresponding to all the turbines in the wind farm. The three signals $\{u(t),\phi(t),P(t)\}$ are sampled at $1$Hz. From now on, these three effective observables for the wind farm are named the {\it wind speed $u(t)$}, the {\it wind direction $\phi(t)$} and the {\it power output $P(t)$}.

\subsection{Ten-minute dynamics}
\label{sec:dyn10min}

\begin{figure}[!h]
   \centering
   \begin{minipage}[t]{0.48\linewidth}
      \centering
      \includegraphics[width=0.8\linewidth]{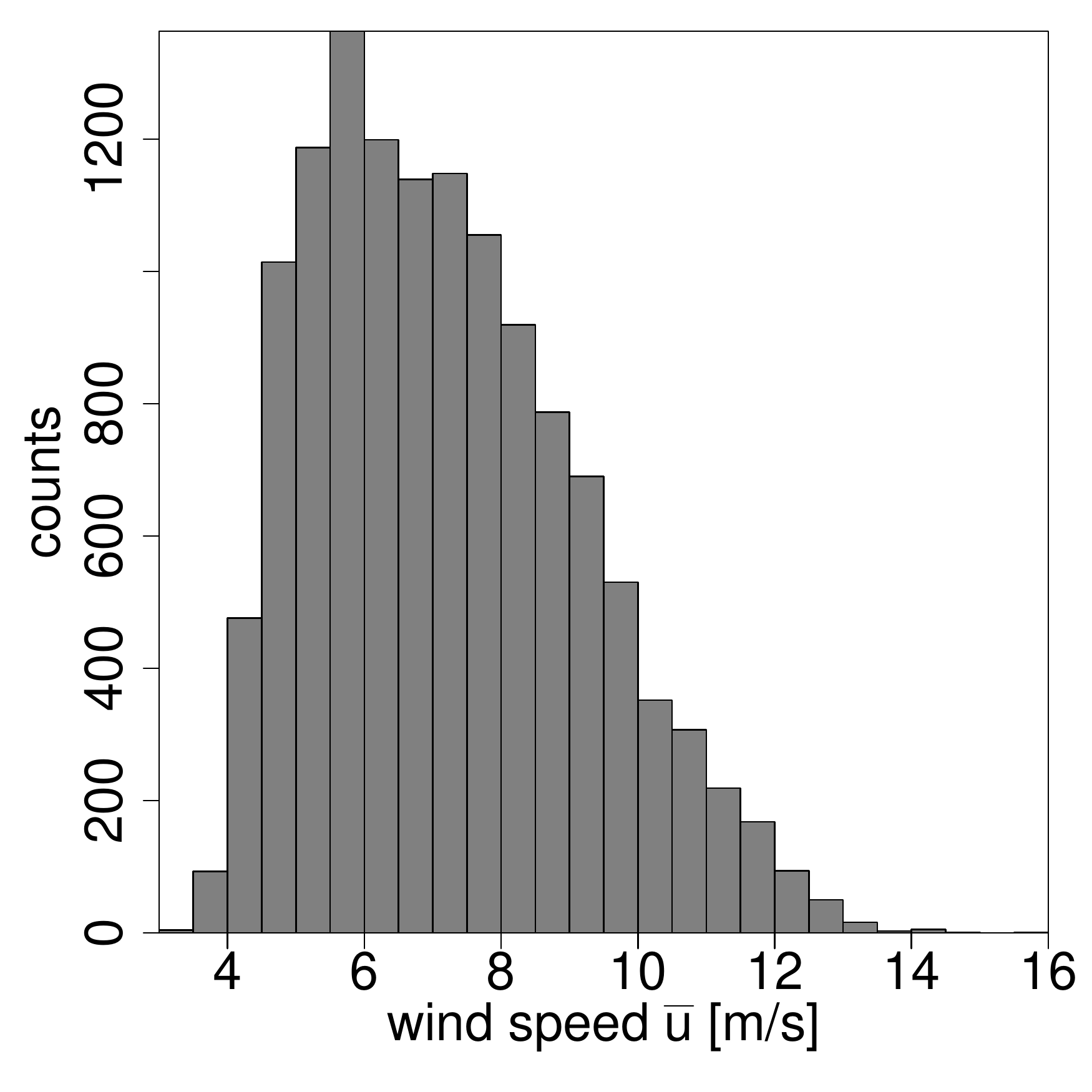}
      \caption{Histogram of ten-minute average wind speed $\bar{u}$ with a resolution $\Delta u=0.5$ m/s.}
      \label{fig:um_hist}
   \end{minipage}%
   \hspace{0.5cm}%
   \begin{minipage}[t]{0.48\linewidth}
      \centering
      \includegraphics[width=0.8\linewidth]{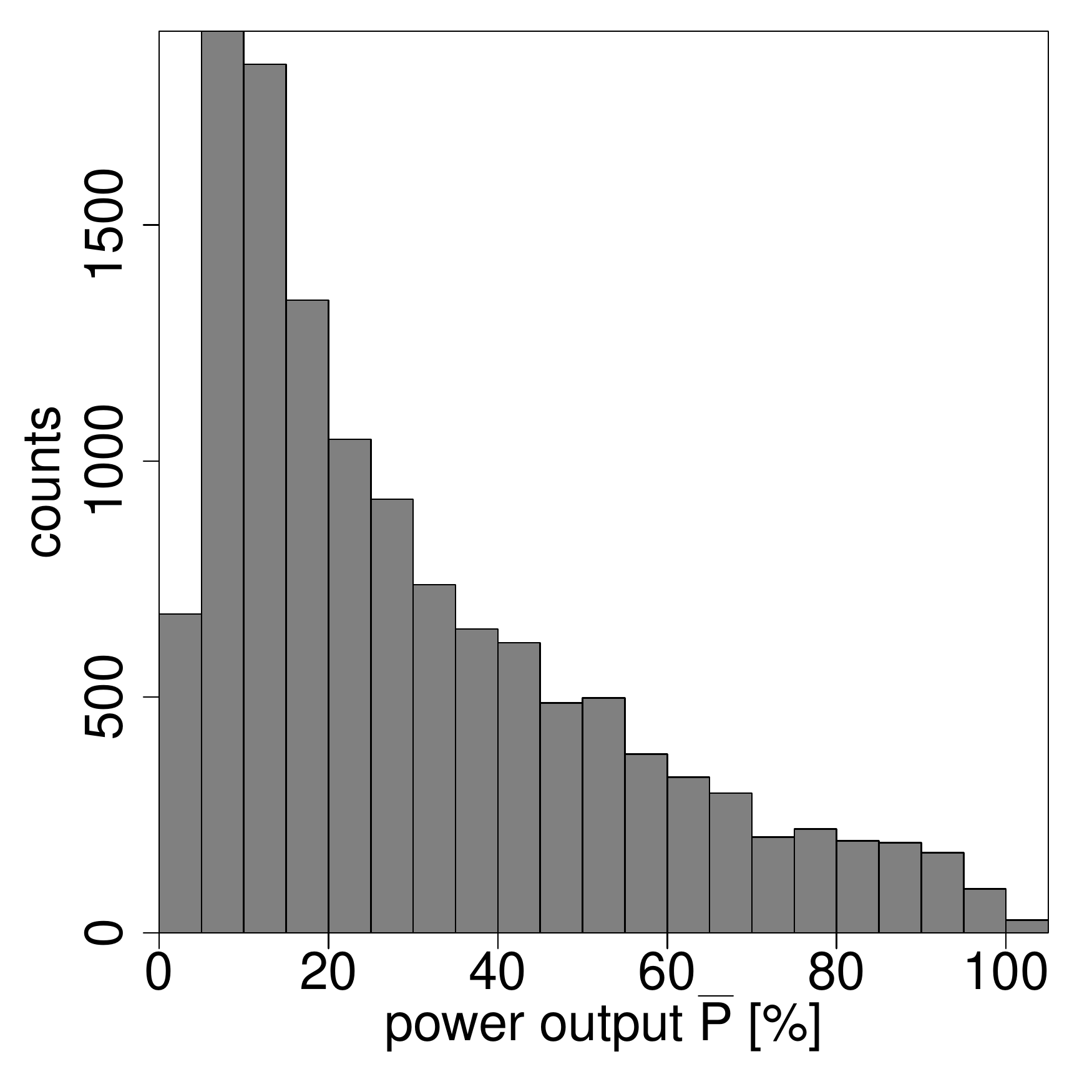}
      \caption{Histogram of ten-minute average power output $\bar{P}$ with a resolution $\Delta P=5$ \%.}
      \label{fig:pm_hist}
   \end{minipage}
\end{figure}

While our analysis focuses on 1-Hz dynamics, we first analyze the wind farm dynamics on the basis of ten-minute averaged results. The ten-minute analysis gives a coarse picture of the wind farm dynamics. Within each ten-minute time period, the average wind speed $\bar{u}$ is calculated from the 600 samples $u(t)$ following
\begin{eqnarray}
\bar{u}=\frac{1}{600} \sum_{t=1}^{600} u(t) \, .
\label{eq:u10min}
\end{eqnarray}
Similarly, the ten-minute average wind direction $\bar{\phi}$ and power output $\bar{P}$ are calculated from $\phi(t)$ and $P(t)$. Histograms are presented in figures \ref{fig:um_hist} and \ref{fig:pm_hist} for $\bar{u}$ and $\bar{P}$. From the $7,775$ ten-minute periods measured, we can observe the Weibull-like shape of the wind speed histogram, as usually observed for atmospheric wind measurements. This is an indication that averaging over all turbine anemometers following equation (\ref{eq:u}) conserves the wind speed histogram. The histogram of power output shows that power values below $20\%$ are most probable, and that the probability of measuring a higher power value decays rapidly. However, it is interesting to note that power values up to $105\%$ of the farm rated power are measured, in which case all turbines deliver slightly above their rated power.

\begin{figure}[!h]
   \centering
   \begin{minipage}[t]{0.48\linewidth}
      \centering
      \includegraphics[width=0.8\linewidth]{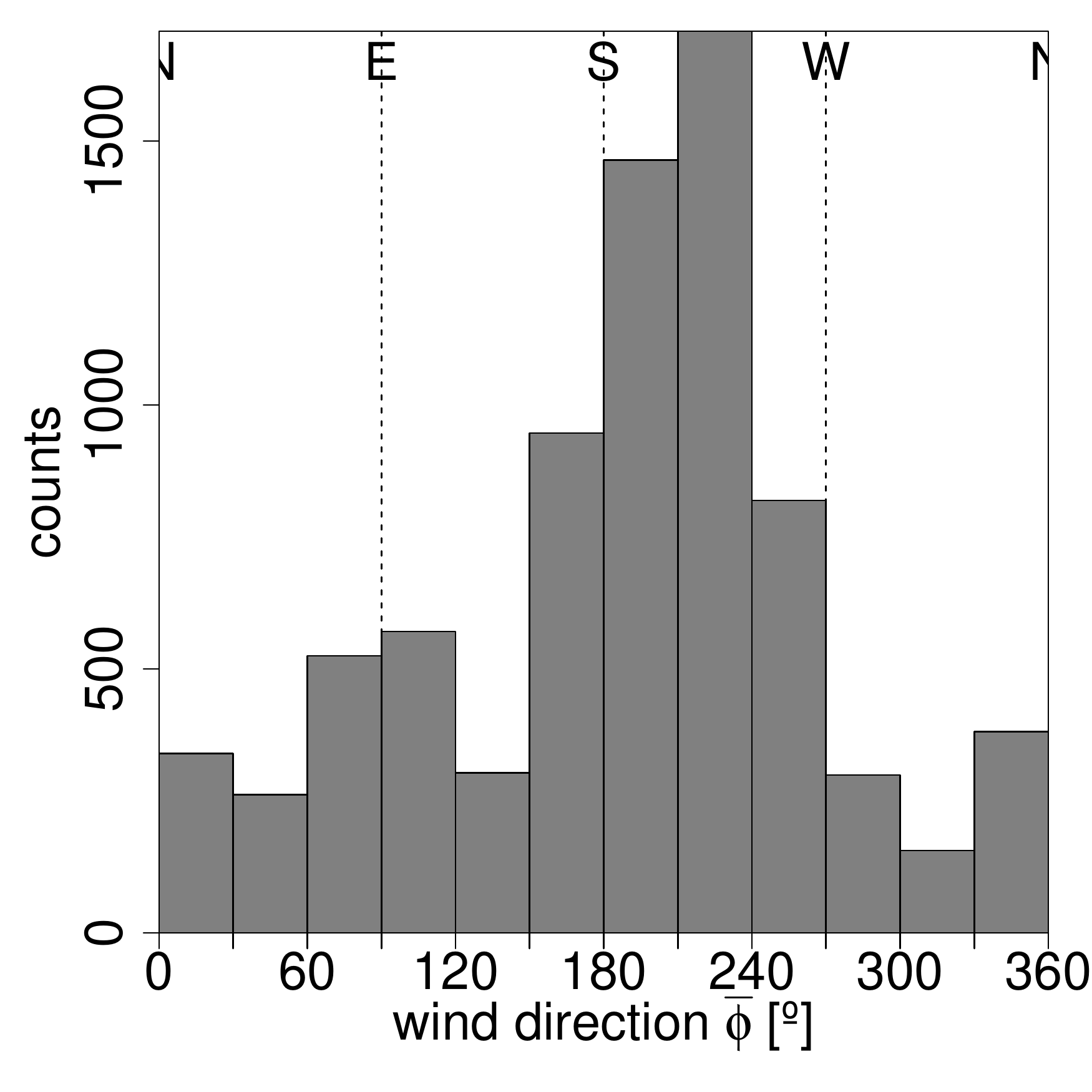}
      \caption{Histogram of ten-minute average wind direction $\bar{\phi}$ with a resolution $\Delta \bar{\phi}=30^{\circ}$. The four cardinal directions are indicated for reference.}
      \label{fig:am_hist}
   \end{minipage}%
   \hspace{0.5cm}%
   \begin{minipage}[t]{0.48\linewidth}
      \centering
      \includegraphics[width=0.8\linewidth]{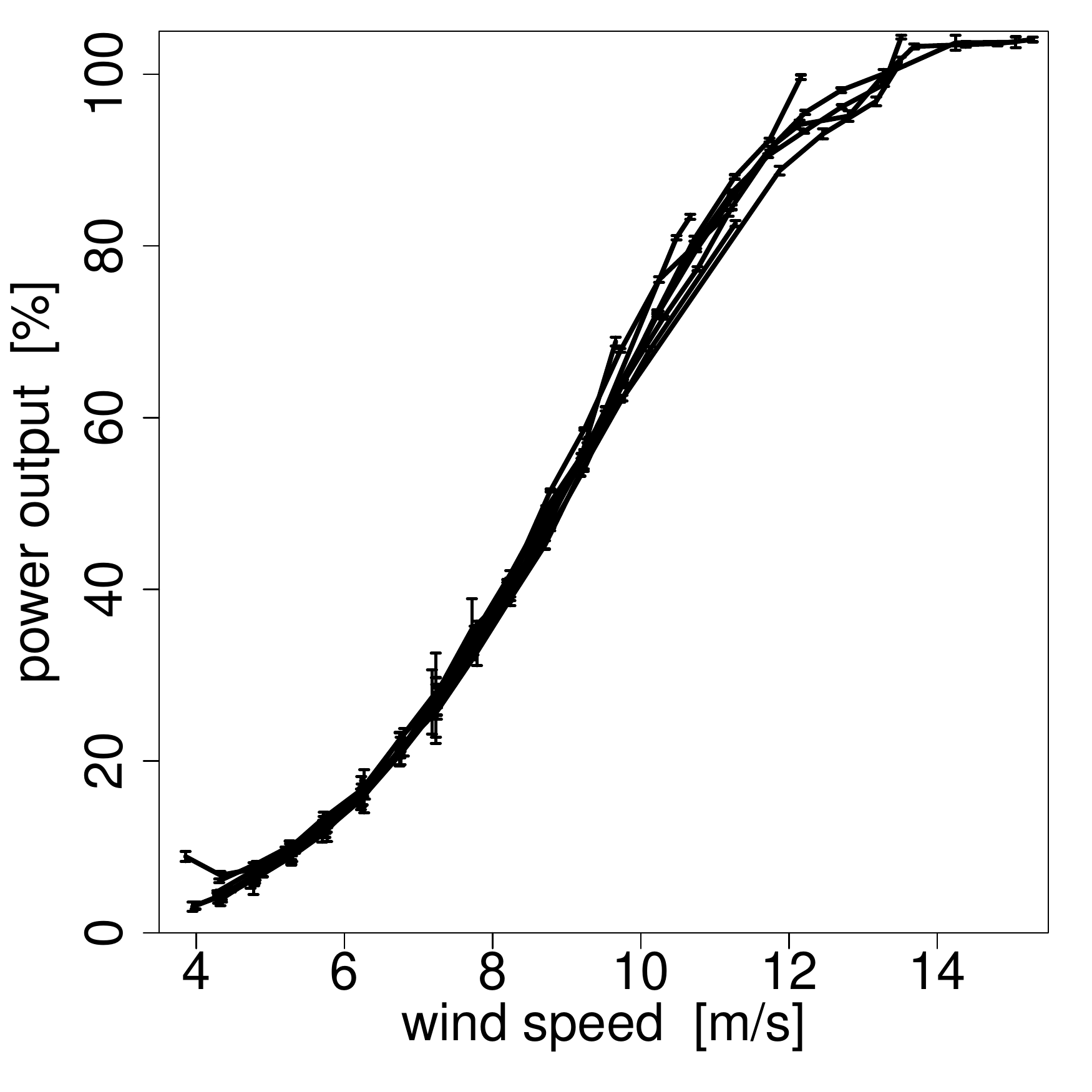}
      \caption{IEC-like power curve $P_{IEC}(\bar{u}_j,\bar{\phi}_k)$ with corresponding error bars for the wind sectors $k\in \{1,12\}$.}
      \label{fig:iec_sectors}
   \end{minipage}
\end{figure}

A histogram of wind directions $\bar{\phi}$ is presented in figure \ref{fig:am_hist} for 12 wind sectors of size $\Delta\bar{\phi}=30^{\circ}$. The most frequent wind directions lie within the South-West sector, as expected from the wind farm location. From now on, the data will be systematically separated for the 12 wind sectors $k$ such that $(k-1) \cdot 30^{\circ} \leq \bar{\phi} < k \cdot 30^{\circ}$.

We define the IEC-like power curve for the wind farm by adapting the norm IEC 61400-12-1 \cite{IEC-12} originally designed for a single wind turbine \footnote{From now on, we refer to the IEC-like power curve as {\it IEC power curve}. We stress that it is not the original procedure, but our own adaptation of the IEC standard to a wind farm.}. For each wind sector $k$ of size $\Delta \bar{\phi}=30^{\circ}$, we sort the ten-minute averages of wind speed $\bar{u}$ and power output $\bar{P}$. We then average the ten-minute averages in each wind speed bin $j$ of size $\Delta \bar{u}=0.5$ m/s. This yields an IEC power curve $P_{IEC}(\bar{u}_j,\bar{\phi}_k)$ for each wind sector, as presented in figure \ref{fig:iec_sectors}. The power curves look similar to the typical IEC power curve of a wind turbine \cite{Burton2001}. A cut-in wind speed is found at around $4$ m/s, and the rated farm power is reached at roughly $13$ m/s, as specified for the single wind turbines in the farm. We observe some deviations in power performance of up to $10\%$ depending on the wind direction. This is due to the various wake effects (reduced wind speed and power downstream of wind turbines) that change with the inflow direction. This implies that the global performance of the farm depends on the inflow direction, and modeling or monitoring applications must be direction-dependent.

\subsection{1Hz dynamics}
\label{sec:dyn1Hz}

\begin{figure}[!h]
   \centering
   \begin{minipage}[t]{0.48\linewidth}
      \centering
      \includegraphics[width=0.8\linewidth]{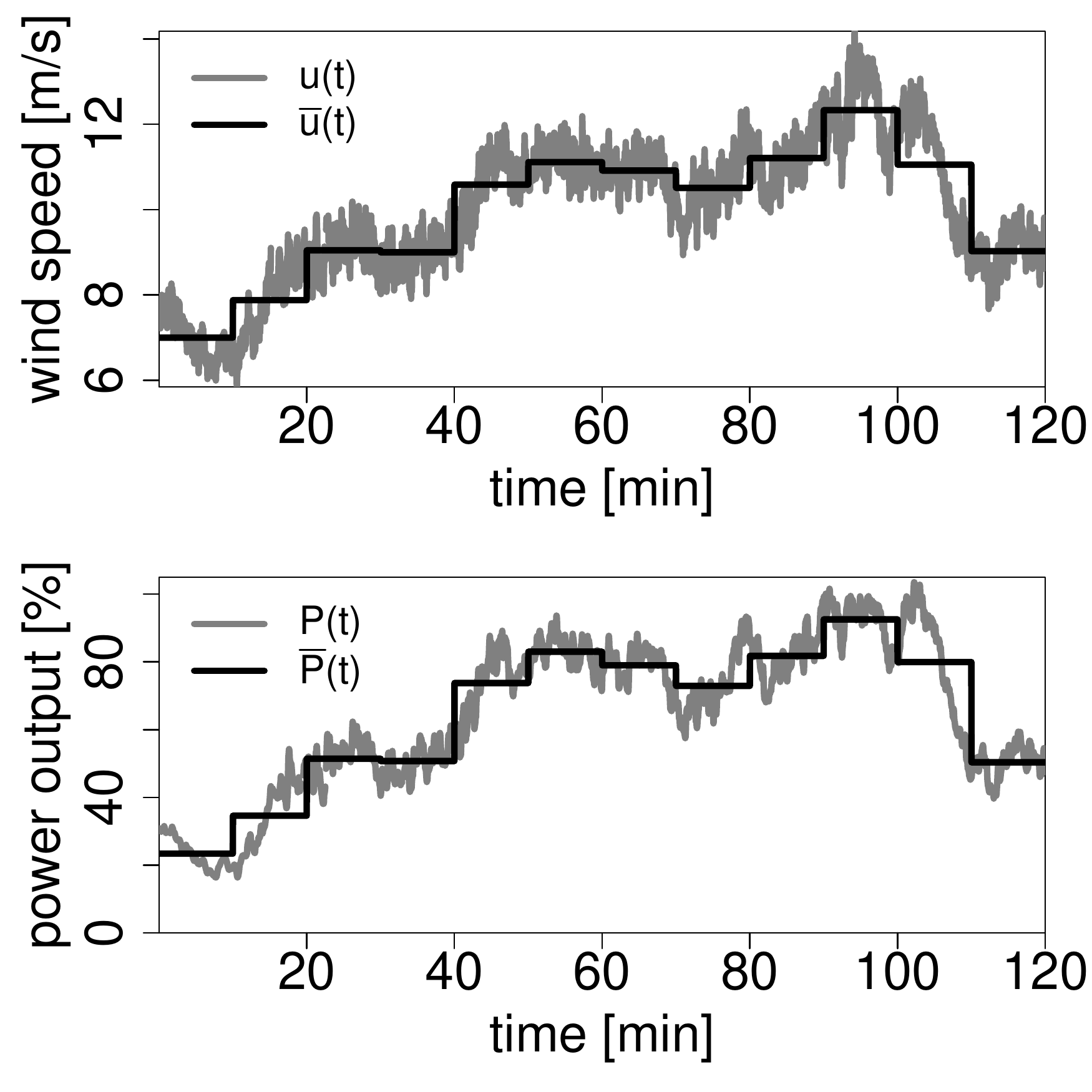}
      \caption{Two-hour excerpts of (upper) wind speed $u(t)$; (lower) power output measurement $P(t)$. The gray lines indicate the $1$Hz signals, and the black lines represent the ten-minute averages $\bar{u}$ and $\bar{P}$. One can note the strong correlation between the wind speed and power output.}
      \label{fig:series8b}
   \end{minipage}%
   \hspace{0.5cm}%
   \begin{minipage}[t]{0.48\linewidth}
      \centering
      \includegraphics[width=0.8\linewidth]{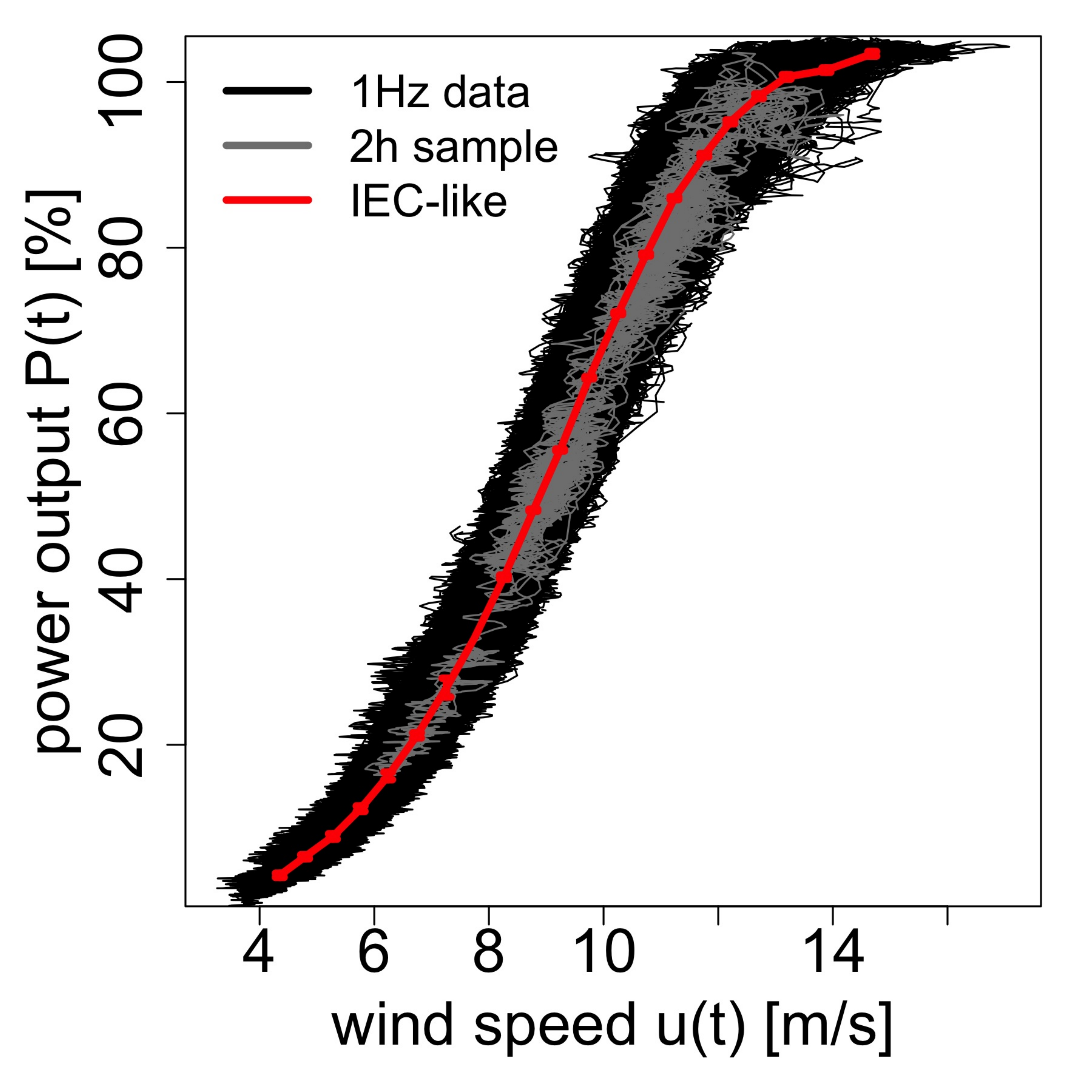}
      \caption{Measured data $\{u(t);P(t)\}$ (black trajectory) at a sampling frequency of $1$Hz for sector $k=8$. The 2-hour excerpt from figure \ref{fig:series8b} is overlaid (gray trajectory). The IEC power curve of figure \ref{fig:iec_sectors} is displayed for this sector (red curve with error bars).}
      \label{fig:uP_1Hz}
   \end{minipage}
\end{figure}

We focus in this paper on wind farm dynamics at $1$Hz. This focus is motivated by observations of measurement data, as presented in figure \ref{fig:series8b}. The effective wind speed $u(t)$ and power output $P(t)$ display fast fluctuations at $1$Hz. While the ten-minute averages $\bar{u}$ and $\bar{P}$ give a rough estimate, the actual dynamics of the wind farm happen on a faster time scale. The wind farm filters the fastest wind fluctuations in time scales of few seconds due to its inertia. However, wind speed changes happening within seconds and minutes are converted into power output changes. The high frequency data also shows wind gusts that are underestimated by ten-minute averages. It is the case in figure \ref{fig:series8b} e.g. at time $100$ min $<$ $t$ $<$ $110$ min, where the wind speed $u(t)$ drops by $5$m/s (while $\bar{u}$ only drops by $2$m/s). In the same time, the power output $P(t)$ drops by $60\%$ (while $\bar{P}$ only drops by $30\%$). This indicates first that rapid and large changes are measured for the wind speed and power output, and second that ten-minutes averages obviously underestimate the amplitude of these rapid changes.

The $\{u,P\}$ space is common for power curve methods because it is the phase space representing the input/output variables $u/P$ for a wind turbine. We use this representation for the wind farm in figure \ref{fig:uP_1Hz}. The fast 1Hz data fluctuates around the IEC power curve with deviations in power output of up to $\pm 20\%$. This shows that large errors up to $20\%$ can be generated by using an IEC power curve for power estimation / modeling. The 2h sample shown in figure \ref{fig:series8b} is overlaid in figure \ref{fig:uP_1Hz}, illustrating the volatility of the conversion process. We also observe that the conversion process $u(t)\to P(t)$ is a highly dynamical process that portrays well how the wind farm functions. We promote the two variables $u(t)$ / $P(t)$ as input / output for the wind farm, and present a stochastic model for the conversion process $(u(t),\bar{\phi})\to P(t)$ in the next section. It should be noted that from now on, the analysis is systematically conditioned on the wind sector $k$ defined by the ten-minute average direction $\bar{\phi}$ \footnote{While we have access to the effective wind direction $\phi(t)$ at $1$Hz, our results show that the ten-minute average $\bar{\phi}$ is a more representative measure. Wake effects are expected to be slow over the farm size of several km, explaining why we favor $\bar{\phi}$ over the fluctuating direction $\phi(t)$.}.

\section{Modeling the conversion process of a wind farm}
\label{sec:model}

We propose in this section a stochastic approach to model the conversion process of the wind farm $(u(t),\bar{\phi})\to P(t)$. We extend a stochastic model originally developed for wind turbines \cite{Milan2011a} to an entire wind farm, based on observations in section \ref{sec:dyn1Hz}. This method converts the wind speed / direction signals into a stochastic signal of power output, after a proper parametrization during a training period.

Three variations of the stochastic method are presented, see table \ref{table:models}. Model 1 (estimated) generates a power output signal $P_1(t)$ at 1Hz using a wind speed signal $u(t)$ at 1Hz. The parameters of model 1 are estimated directly during the training period, see subsection \ref{sec:model1}. Model 2 (parametric) functions similarly to model 1 except that the parameters are no longer those estimated from the training period but a parametric version of them, see subsection \ref{sec:model2}. Model 3 (parametric 10-min) extends model 2 by including a pre-model that first models a 1Hz wind speed signal from ten-minute wind data, then follows instructions from model 2, see subsection \ref{sec:model3}. Model 3 can be used when only ten-minute wind data is available. Additionally, a fourth model based on the IEC power curve is used to compare to the stochastic models 1-3. All power output signals (one measured and four modeled) are compared in subsection \ref{sec:model_results}.
\begin{table}[!h]
\centering
\begin{tabular}{  l || c | c || c  }
\hline
\multicolumn{1}{c||}{\multirow{2}{*}{\textbf{SIGNAL}}}	& \multicolumn{2}{|c||}{\textbf{INPUT}}			& \textbf{OUTPUT}		\\ \cline{2-4}
											& \textbf{wind speed}		& \textbf{wind direction}		& \textbf{power output}	\\ \hline \hline
\textbf{reference (measurement)}			& measured $u(t)$			& measured 10-min $\bar{\phi}$	& measured $P(t)$		\\ \hline
\textbf{model 1 (estimated)}				& measured $u(t)$			& measured 10-min $\bar{\phi}$	& modeled $P_1(t)$		\\ \hline
\textbf{model 2 (parametric)}				& measured $u(t)$			& measured 10-min $\bar{\phi}$	& modeled $P_2(t)$		\\ \hline
\textbf{model 3 (parametric+10-min wind)}	& measured 10-min $\bar{u}$	& measured 10-min $\bar{\phi}$	& modeled $P_3(t)$		\\ \hline
\textbf{IEC power curve}					& measured $u(t)$			& measured 10-min $\bar{\phi}$	& modeled $P_{IEC}(t)$	\\ \hline
\end{tabular}
\caption{Overview of the different models presented. Each model converts the wind speed and wind direction input signals into an output power signal.}
\label{table:models}
\end{table}

\subsection{Stochastic estimation of the conversion process}
\label{sec:model1}
In this subsection, the model parameters are estimated directly from data measured during the training period. This approach is the direct extension of the wind turbine model presented in \cite{Milan2011a}. We observed in figure \ref{fig:uP_1Hz} that the 1Hz trajectory in $\{u,P\}$ space fluctuates around the IEC power curve $P_{IEC}(\bar{u},\bar{\phi})$. Simply put, turbulent wind fluctuations drive the farm dynamics away from the power curve, that represents the average behavior aimed by the wind farm. Ref. \cite{Anahua2007,Gottschall2008} recognized this property for single wind turbines and proposed to describe these dynamics with a relaxation model towards the power curve. The power curve is simply seen as an attractor, around which the system fluctuates.

Such dynamics can be reproduced by a stochastic drift / diffusion model. In this paradigm, the attractor towards some hypothetical power curve $P(u)$ is represented by a drift coefficient $D^{(1)}$, while a diffusion coefficient $D^{(2)}$ models additional fluctuations. These two coefficients (also called Kramers-Moyal coefficients in stochastic theory) can be estimated directly from measured data \cite{Risken1996,Friedrich2011} collected during the training period. They describe the dynamics of the power output $P(t)$ conditioned on the 1Hz wind speed $u(t)$ and the ten-minute average wind direction $\bar{\phi}$. The system evolves in a three-dimensional space with variables $P(t)$, $u(t)$ and $\bar{\phi}$. The drift and diffusion coefficients must then be defined as the three-dimensional arrays
\begin{eqnarray}
D^{(n)}(P(t),u(t),\bar{\phi})	&=& \frac{1}{n!} \lim_{\tau \to 0} \frac{1}{\tau}M^{(n)}(P(t),u(t),\bar{\phi},\tau) \\
								&=& \frac{1}{n!} \frac{\partial M^{(n)}(P(t),u(t),\bar{\phi},\tau)}{\partial \tau}\Bigg|_{\tau=0}  \, ,
\label{eq:kmc}
\end{eqnarray}
with $n=1$ for the drift coefficient and $n=2$ for the diffusion coefficient. As shown in Ref. \cite{Boettcher2006}, they are the partial derivatives with respect to the time increment $\tau$ of the conditional moments\footnote{This definition of the coefficients based on a derivation allows for a more robust estimation in the presence of measurement noise that spoils the data.}
\begin{eqnarray}
M^{(n)}(P(t),u(t),\bar{\phi},\tau)	&=& \int\limits_{-\infty}^{\infty} \big[\,P(t+\tau)-P(t) \,\big]^n \,\,\, f(P(t+\tau)|P(t),u(t),\bar{\phi}) \,\,\, dP(t+\tau) \\
&=& \Big \langle \,\big[\,P(t+\tau)-P(t) \,\big]^n\,\, \Big|\,\,P(t) \, ,\, u(t),\, \bar{\phi}\, \Big \rangle \, ,
\label{eq:cond_mom}
\end{eqnarray}
where the operator $\langle A|B \rangle$ represents the conditional mean of $A$ for condition $B$. The conditional moment $M^{(n)}(P(t),u(t),\bar{\phi},\tau)$ is the $n$-th moment of the conditional probability $f(P(t+\tau)|P(t),u(t),\bar{\phi})$ of the power increment $P(t+\tau)-P(t)$ conditioned on the values of $P(t)$, $u(t)$ and $\bar{\phi}$.

It is not possible to calculate the derivative in equation (\ref{eq:kmc}) exactly from measured data. The time increment $\tau$ is limited to the smallest value min$(\tau)=1/f_s$, where $f_s$ is the sampling frequency. In our case, min$(\tau)=1$s, and the limit $\tau \to 0$ cannot be performed. However, a Taylor expansion of the conditional moments gives in first-order approximation
\begin{eqnarray}
M^{(n)}(*,\tau)	&=& M^{(n)}(*,\tau=0) + \tau \cdot \frac{\partial M^{(n)}(*,\tau)}{\partial \tau}\Bigg|_{\tau=0} + o(\tau^2)\\
				&=& \tau \cdot \frac{\partial M^{(n)}(*,\tau)}{\partial \tau}\Bigg|_{\tau=0} + o(\tau^2)\\
				&=& n! \tau \cdot D^{(n)}(*) + o(\tau^2) \, ,
\label{eq:cond_mom_taylor}
\end{eqnarray}
where $*$ stands for the variables $P(t)$, $u(t)$ and $\bar{\phi}$. The relation $M^{(n)}(*,\tau=0)=0$ is obvious from equation (\ref{eq:cond_mom}). For small non-zero $\tau$ values, the derivative in equation (\ref{eq:kmc}) becomes to a first-order approximation
\begin{eqnarray}
D^{(n)}(P(t),u(t),\bar{\phi}) \simeq \frac{M^{(n)}(P(t),u(t),\bar{\phi},\tau)}{n! \tau} \, .
\label{eq:kmc2}
\end{eqnarray}
Additionally, the discrete nature of measured data has further implications on the applicability of theories developed for continuous stochastic processes. It was observed by \cite{Sura2002,Gottschall2008b} that when estimating the coefficients from a dataset that is not sampled with a high enough sampling frequency, an artificial parabolic term is added to the diffusion coefficient. A simple ansatz exists in order to remove this artifact from $D^{(2)}$, where equation (\ref{eq:kmc2}) is used with a second conditional moment $M^{(2)}$ modified as \cite{Friedrich2011}
\begin{eqnarray}
M^{(2)}(P(t),u(t),\bar{\phi},\tau)	= \Big \langle \,\big[\,P(t+\tau)-P(t) - \tau \, D^{(1)}(P(t),u(t),\bar{\phi}) \,\big]^2 \,\, \Big|\,\,P(t) \, ,\, u(t),\, \bar{\phi}\, \Big \rangle \, .
\label{eq:cond_mom_finite_sampling}
\end{eqnarray}
This approach is used systematically in this paper in order to minimize finite sampling artifacts.

Concretely speaking, the three measurement signals $P(t)$, $u(t)$ and $\bar{\phi}$ are collected during the training period and sorted in the three-dimensional phase space $\{P,u,\bar{\phi}\}$. Each axis, e.g. the P-axis is split into $N_P$ equidistant intervals of size $\Delta P$. The phase space is then cut into $N_P \times N_u \times N_{\bar{\phi}}$ bins (small cubes) of size $\{\Delta P,\Delta u,\Delta \bar{\phi}\}$. Let us consider one bin $(i,j,k)$ which contains all the data samples that satisfy $P(t) \in (P_i \pm \Delta P/2$), $u(t) \in (u_j \pm \Delta u/2$) and $\bar{\phi} \in (\bar{\phi}_k \pm \Delta \bar{\phi}/2$). From all these data samples, the conditional moments $M^{(n)}(P_i,u_j,\bar{\phi}_k,\tau)$ are calculated  following equation (\ref{eq:cond_mom}). Then the coefficients $D^{(n)}(P_i,u_j,\bar{\phi}_k)$ can be estimated from the conditional moments following equation (\ref{eq:kmc2}). This operation is then repeated for each of the $N_P \times N_u \times N_{\bar{\phi}}$ bins, such that the drift and diffusion coefficients can be estimated in the entire three-dimensional phase space \footnote{The estimation is only possible in regions of the phase space that were visited during the training period, i.e. sufficient events are in this bin so that the corresponding mean values can be estimated. The training period should then be long enough for the measurement to span all the regions of interest.}. Here we choose $\Delta u=0.5$ m/s, $\Delta \bar{\phi}=30^{o}$ and $\Delta P=P_r/50=2\%$.

The binning procedure allows to characterize the dynamics locally in the phase space. The information is averaged following equation (\ref{eq:cond_mom}) in each local bin (rather than in time like the IEC procedure). This means that in each bin of the power-speed-direction space, the power dynamics are described by the drift and diffusion coefficients. In bin $(i,j,k)$, the drift coefficient $D^{(1)}(P_i,u_j,\bar{\phi}_k)$ represents the first moment, i.e. the mean value of the power change $\frac{dP(t)}{dt}=\lim_{\tau \to 0} \frac{P(t+\tau)-P(t)}{\tau}$ calculated from all data samples in the bin. Similarly, the diffusion coefficient $D^{(2)}(P_i,u_j,\bar{\phi}_k)$ represents the variance of the power change. In summary, $D^{(1)}$ and $D^{(2)}$ represent the mean and variance of the power change $\frac{dP(t)}{dt}$. They quantify how much the power changes on average (drift), and how much deviation is expected from that average change (diffusion). Although not mathematically exact, one could conceptualize the coefficients as
\begin{eqnarray}
D^{(1)}(P_i,u_j,\bar{\phi}_k)	&\sim& \Bigg \langle \frac{dP(t)}{dt} \Bigg|\,\,P_i,u_j,\bar{\phi}_k\, \Bigg \rangle \, , \\
D^{(2)}(P_i,u_j,\bar{\phi}_k)	&\sim& \Bigg \langle \Bigg(\frac{dP(t)}{dt}\Bigg)^2 dt \Bigg|\,\,P_i,u_j,\bar{\phi}_k\, \Bigg \rangle \, .
\label{eq:kmc3}
\end{eqnarray}

\begin{figure}[!h]
   \centering
   \begin{minipage}[t]{0.48\linewidth}
      \centering
   	 \includegraphics[width=\columnwidth]{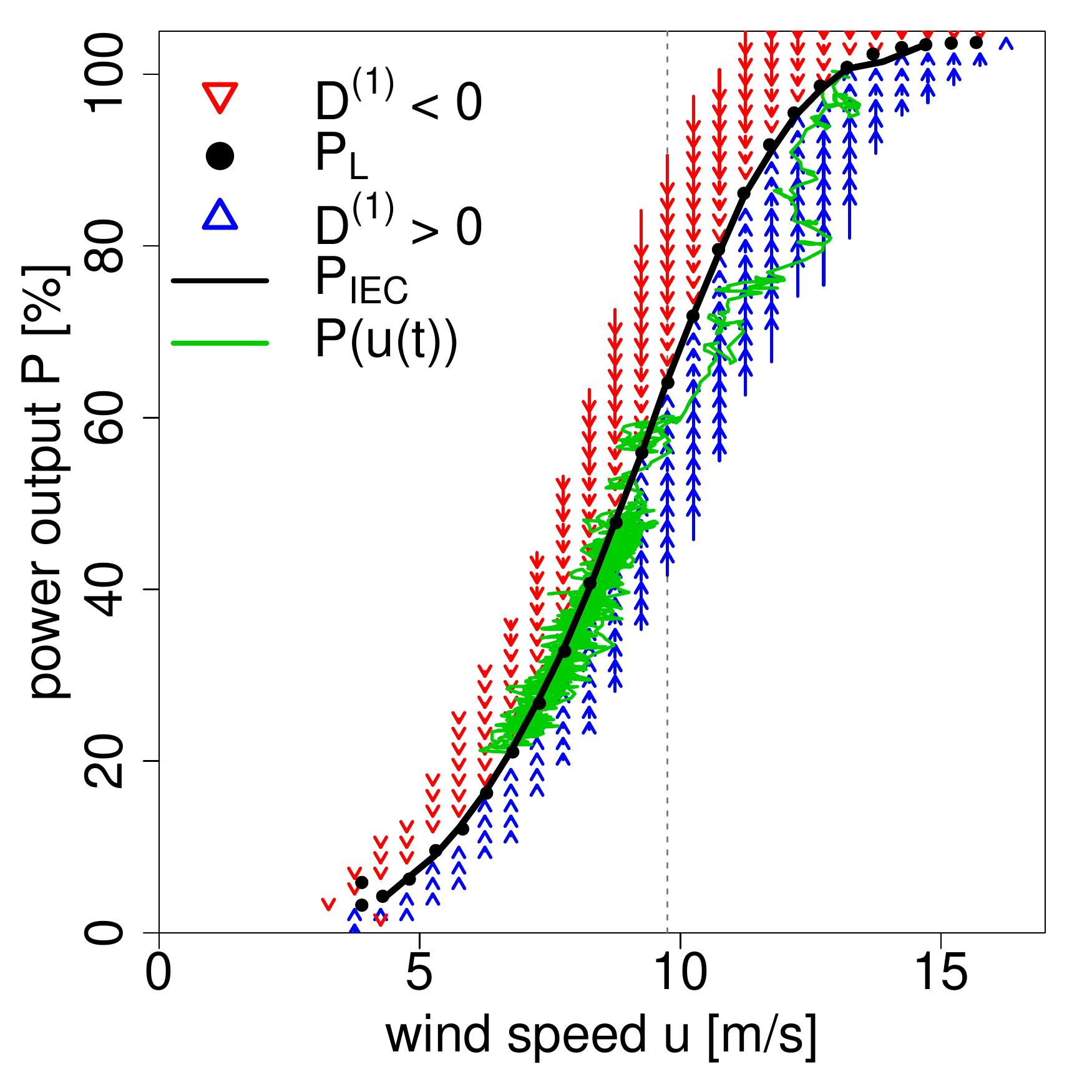}  
	\caption{drift coefficient $D^{(1)}(P_i,u_j;\bar{\phi}_k)$ estimated from equation (\ref{eq:kmc2}) for the sector $k=8$ ($\bar{\phi}\simeq 225^\circ$). The drift is positive (blue arrows upwards) where the power output increases, and negative (red arrows downwards) where the power decreases. Its stable fixed points define the Langevin power curve $P_L(u_j,\bar{\phi}_k)$ (black dots), that matches the IEC power curve $P_{IEC}(u_j,\bar{\phi}_k)$ (black line). A 1Hz trajectory is shown (green line). The wind speed bin $j=20$ ($u\simeq 9.75$m/s) is indicated (dashed gray line) corresponding to figure \ref{fig:drift_potential}.}
  \label{fig:drift}
   \end{minipage}%
   \hspace{0.5cm}%
   \begin{minipage}[t]{0.48\linewidth}
      \centering
   	 \includegraphics[width=\columnwidth]{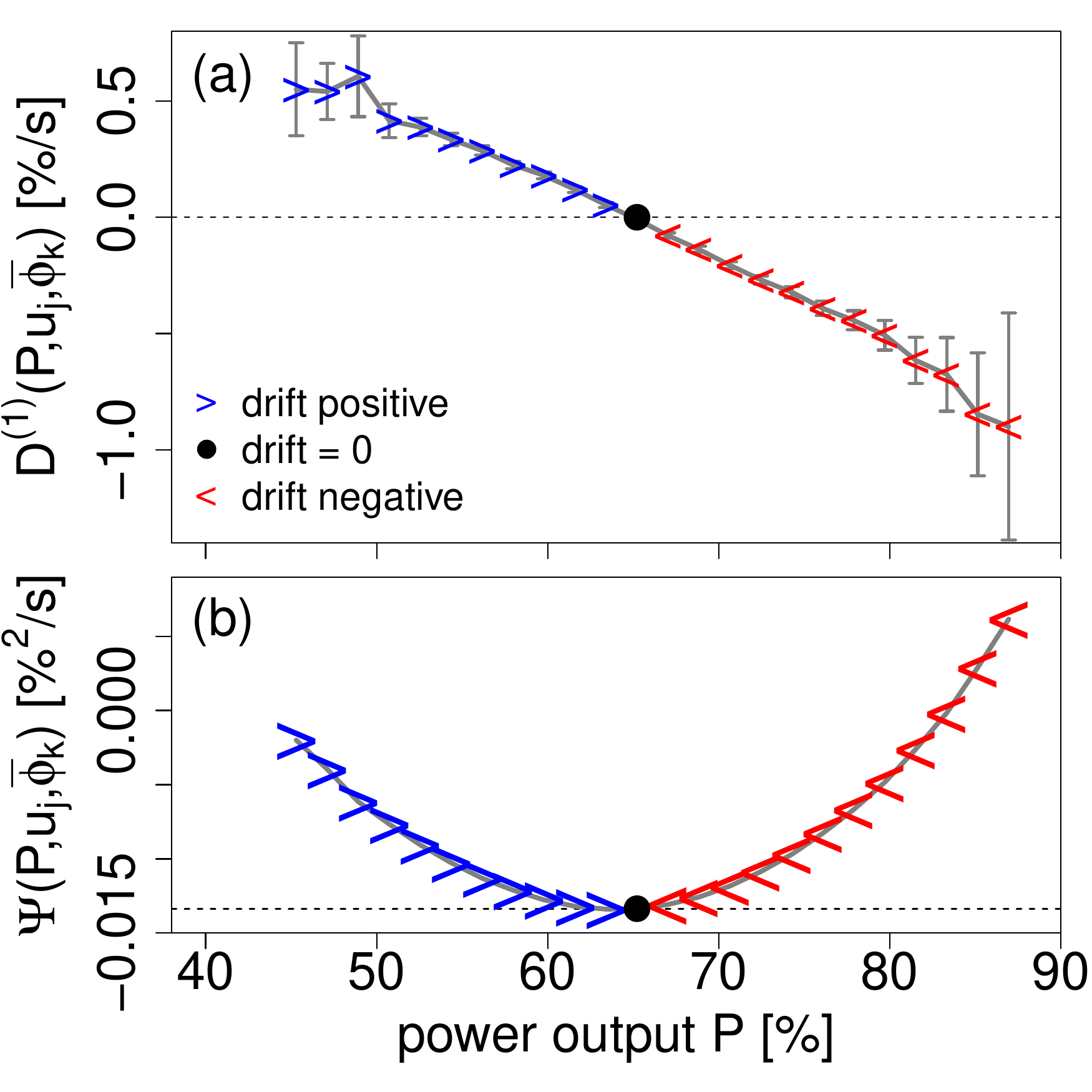}  
	\caption{(a) drift coefficient $D^{(1)}(P,u_j;\bar{\phi}_k)$; (b) drift potential $\Psi(P,u_j;\bar{\phi}_k)$ for wind speed bin $j=20$ ($u\simeq 9.75$m/s) and sector $k=8$ ($\bar{\phi}\simeq 225^\circ$) (gray line with error bars in case of the drift coefficient). Symbols are superimposed for positive drift (blue arrows), negative drift (red arrows) and zero drift values (black dot). The zero drift and minimum potential values are indicated (horizontal dashed line), corresponding to the value of the Langevin power curve $P_L(u_j;\bar{\phi}_k)$.}
  \label{fig:drift_potential}
   \end{minipage}
\end{figure}

The drift coefficient is presented in figure \ref{fig:drift}. In bin $(i,j,k)$, if $D^{(1)}>0$ the drift is represented by a blue arrow pointing up because a positive drift means a power increase. Similarly, if $D^{(1)}<0$ the drift is represented by a red arrow pointing down describing a power decrease. The dynamics of the wind farm appear clearly. In all bins located below the power curve, the drift is positive and the power increases (on average). Analogously, in all bins above the power curve, the power decreases as the drift coefficient is negative. A clear drift towards the power curve appears. The drift coefficient is a map of the conversion process, that extends the information of the ten-minute averaged IEC power curve to 1Hz dynamics. A trajectory is overlaid to the drift coefficient in figure \ref{fig:drift} to illustrate the concept of a drift towards the attractive power curve. The drift coefficient is shown in figure \ref{fig:drift_potential}(a) for one wind speed / direction bin and confirms what was observed in figure \ref{fig:drift}.

One can also define the potential of the drift coefficient as
\begin{eqnarray}
\Psi(P,u_j,\bar{\phi}_k) = - \int_{-\infty}^P D^{(1)}(P',u_j,\bar{\phi}_k) \, dP' \, .
\label{eq:psi}
\end{eqnarray}
The potential is a more intuitive representation of the dynamics, see figure \ref{fig:drift_potential}(b). The system tends to minimize its potential by drifting towards an attractive fixed point (that corresponds to a local minimum of the potential). Around the attractive fixed point, the potential climbs to indicate the repulsion of the system to move away from the power curve. However, the system keeps following the non-stationary wind condition (changing $j$, $k$ bins) and always drifts towards a new fixed point.

Beyond the drift coefficient, the power values towards which the system drifts are of interest. We define this set of attractive fixed points as a {\it Langevin power curve} $P_L(u_j,\bar{\phi}_k)$ following
\begin{eqnarray}
P_L(u_j,\bar{\phi}_k)\Leftrightarrow
\begin{cases}
        D^{(1)}(P_L(*),*)=0 \\
        \Big( \frac{\partial D^{(1)}(P,*)}{\partial P} \Big)_{P=P_L(*)} <0
\end{cases}
\label{eq:lpc}
\end{eqnarray}
where $*$ stands for $(u_j,\bar{\phi}_k)$. At a fixed point, the drift coefficient is zero. For the fixed point to be attractive, the drift must be decreasing (negative derivative with respect to P). On the contrary, a fixed point with increasing drift would be repulsive. The system drifts towards the Langevin power curve $P_L(u_j,\bar{\phi}_k)$, that represents the basin of attraction for the wind speed $u_j$ and wind sector $\bar{\phi}_k$. The Langevin and IEC power curves have similar values for the data set measured here, see figure \ref{fig:drift}.

The dynamics can be seen as a random-like trajectory circulating around the power curve. This is the signature of a process that contains both deterministic and stochastic dynamics. The true reason for the random nature of the process is not obvious, as wind turbines have deterministic control strategies. At this point, it is important to note that the power dynamics are driven by the wind speed dynamics. In an idealized case with a spatially homogeneous, laminar wind inflow, the system should react in a deterministic manner. If this spatially homogeneous flow would see its speed change everywhere, the drift field should describe the corresponding change in power output satisfyingly. However, this ideal case is not realistic because the wind inflow is turbulent, and contains spatial fluctuations. The wind farm does not react to the effective wind speed $u(t)$ averaged over the whole area, but to the local wind fluctuations acting on the local blade elements of all the rotors, so to say the wind field $\vec{U}(\vec{x},t)$. The effective wind speed $u(t)$ is a macroscopic estimate of the many microscopic influences driving the wind farm (analogously to temperature being a macroscopic estimate of thermodynamic fluctuations). When projecting the system dynamics over this macroscopic variable, the microscopic degrees of freedom become stochastic fluctuations.

A macroscopic variable driven by many microscopic interactions can often be approximated by a low-dimensional Langevin process \footnote{It is a direct analogy to the motion of a macroscopic particle within a flow, that was described by Robert Brown in 1827 as {\it Brownian motion}, or a consequence of synergetic \cite{Haken1983}.}. We can write a Langevin equation that describes the time evolution of the power output $P(t)$ following
\begin{eqnarray}
\frac{dP(t)}{dt}=D^{(1)}(P(t),u(t),\bar{\phi})+\sqrt{D^{(2)}(P(t),u(t),\bar{\phi})}\cdot \Gamma(t) \, ,
\label{eq:langevin}
\end{eqnarray}
where $\Gamma(t)$ is a Gaussian-distributed noise that is uncorrelated, i.e. $\langle \Gamma(t) \Gamma(t') \rangle=2\delta(t-t')$ and with a mean value $\langle \Gamma(t) \rangle=0$. The Langevin equation can be integrated numerically in the It{\^o} sense \cite{Risken1996}
\begin{eqnarray}
P(t+\Delta t)=P(t)+\Delta t \cdot D^{(1)}(P(t),u(t),\bar{\phi})+\sqrt{\Delta t \cdot D^{(2)}(P(t),u(t),\bar{\phi})}\cdot \eta(t) \, 
\label{eq:langevin_euler}
\end{eqnarray}
where $\eta(t)$ is a Gaussian-distributed random variable with mean value $\langle \eta(t) \rangle=0$ and variance $\langle \eta(t)^2 \rangle=2$. \footnote{The noise term $\Gamma(t)$ fluctuates much faster than $\sqrt{D^{(2)}(P(t),u(t),\bar{\phi})}$, that is considered constant during the integration step $\Delta t$. When integrating the stochastic Langevin equation following It{\^o} definition, $\int_t^{t+\Delta t} \sqrt{D^{(2)}(P(t'),u(t'),\bar{\phi})}\cdot \Gamma(t') dt'=\sqrt{D^{(2)}(P(t),u(t),\bar{\phi})} \int_t^{t+\Delta t} \Gamma(t') dt'$. Ref. \cite{Friedrich2011} shows that $\int_t^{t+\Delta t} \Gamma(t') dt'=\sqrt{\Delta t} \cdot \eta(t)$.} The assumption of Gaussianity can be quantified by higher-order Kramers-Moyal coefficients being zero, i.e. $D^{(n)}=0$ for $n>2$ following Pawula theorem \cite{Risken1996}.

The structure of the Langevin equation indicates that the power output $P(t)$ drifts towards the power curve following $D^{(1)}$, but stochastic fluctuations are superimposed using the stochastic noise $\Gamma(t)$ amplified by the diffusion coefficient $D^{(2)}$. This way the wind farm dynamics captured by the drift / diffusion coefficients can be remodeled by solving the Langevin equation. Given any new wind speed signal $u(t)$ and wind direction $\bar{\phi}$ as input, equation (\ref{eq:langevin_euler}) generates a synthetic power output signal $P_1(t)$. This serves as a model for the conversion process ($u(t),\bar{\phi}) \to P_1(t)$. Some results are presented in subsection \ref{sec:model_results} in comparison with other models.

\subsection{Parametric form of the stochastic model}
\label{sec:model2}

\begin{figure}[!h]
  	\begin{center}
   	 \includegraphics[width=\columnwidth]{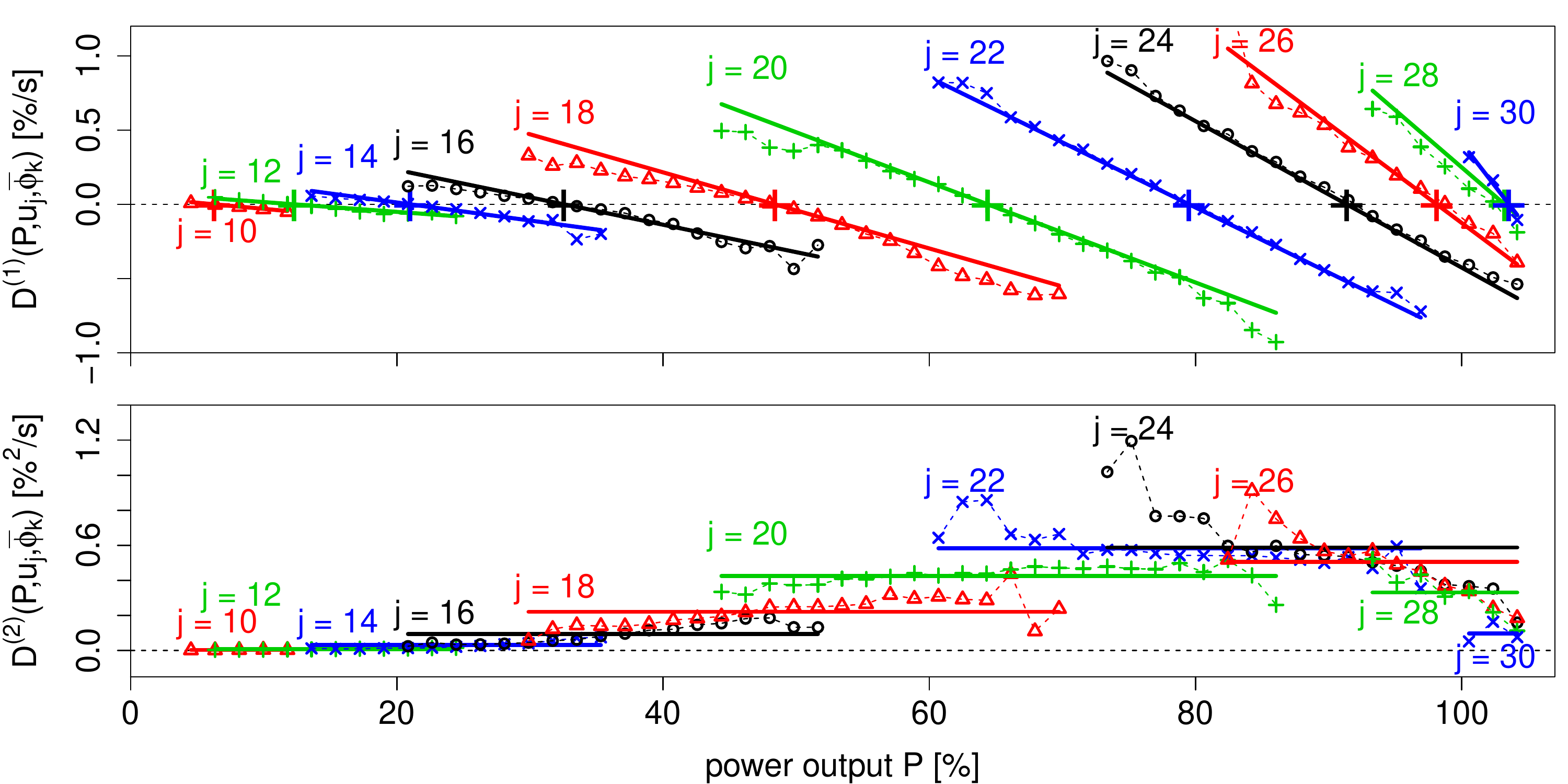}  
	\end{center}
	\caption{(upper) drift coefficient $D^{(1)}(P,u_j,\bar{\phi}_k)$ estimated (symbols and dashed line) for various wind speed bins $j$ (various colors) and sector $k=8$ ($\bar{\phi}\simeq 225^\circ$). The power curve value $P_{IEC}(u_j,\bar{\phi}_k)$ is indicated (colored symbol $+$). The linear fit proposed in equation (\ref{eq:D1_parametric}) is also drawn (bold colored line); (lower) similarly for the diffusion coefficient $D^{(2)}(P,u_j,\bar{\phi}_k)$. The mean value $\beta_{jk}$ of the diffusion coefficient is represented (colored bold line) following equation (\ref{eq:D2_parametric}).
	}
  \label{fig:KMC_parametric}
\end{figure}

The drift and diffusion coefficients $D^{(n)}(P_i,u_j;\bar{\phi}_k)$ estimated from equation (\ref{eq:kmc2}) describe the dynamics of the conversion process. A parametric simplification of the coefficients is presented here in an effort to reduce the number of model parameters, see figure \ref{fig:KMC_parametric}.

We observe that for a given wind speed bin $u_j$ and direction bin $\bar{\phi}_k$, the drift coefficient $D^{(1)}(P,u_j,\bar{\phi}_k)$ is a tolerably linear function of $P$. The stable fixed points are given by the Langevin power curve $P_{L}(u_j,\bar{\phi}_k)$,\footnote{As observed in figure \ref{fig:drift}, the IEC power curve $P_{IEC}(u_j,\bar{\phi}_k)$ is very close to the Langevin power curve $P_{L}(u_j,\bar{\phi}_k)$ for the data set measured. For this reason, the IEC power curve could be used instead of the Langevin power curve in equation (\ref{eq:D1_parametric}) if necessary.} so that the drift coefficient can be simplified as
\begin{eqnarray}
D^{(1)}(P,u_j,\bar{\phi}_k)	\simeq \alpha_{jk} \cdot \big( P - P_{L}(u_j,\bar{\phi}_k) \big) \, ,
\label{eq:D1_parametric}
\end{eqnarray}
where $\alpha_{jk}$ is its slope \footnote{The condition $\alpha_{jk}<0$ must be fulfilled for the model to be stable. This condition corresponds to a converging drift that drives the dynamics towards a stable attractor, the Langevin power curve.}. The drift slope $\alpha_{jk}$ is a direct measure of how fast the wind farm can follow wind fluctuations. A more negative slope (steeper drift) corresponds to a faster reaction of the wind farm \footnote{For example, when the wind speed is around $12$ m/s, $\alpha_{jk}\simeq-0.05$/s. If the power output $P$ is $20\%$ lower than the ideal value $P_{L}$, equation (\ref{eq:D1_parametric}) indicates that $D^{(1)} \simeq \alpha_{jk} \cdot ( P - P_{L} ) \simeq 1\%/$s. When in such conditions, the power output changes on average by $\sim 1\%$ each second.}.

Similarly for the diffusion term $D^{(2)}(P,u_j,\bar{\phi}_k)$, some parametrization can be performed. As illustrated in figure \ref{fig:KMC_parametric}, the diffusion coefficient is almost constant, such that
\begin{eqnarray}
D^{(2)}(P,u_j,\bar{\phi}_k)	\simeq \beta_{jk} = \frac{1}{N_P} \sum_{i=1}^{N_P} D^{(2)}(P_i,u_j,\bar{\phi}_k) \, ,
\label{eq:D2_parametric}
\end{eqnarray}
where the constant $\beta_{jk}$ is the diffusion coefficient averaged over the $N_P$ power bins. The mean diffusion term $\beta_{jk}$ quantifies the random nature of the power fluctuations. Higher values indicate stronger power fluctuations \footnote{Although $\beta_{jk}$ quantifies the intensity of the diffusion coefficient $D^{(2)}(P,u_j,\bar{\phi}_k)$, a measure for the variance of the fluctuations is the ratio $\beta_{jk}/\alpha_{jk}$. That is, the competition between diffusion and drift determines the range of expected fluctuations, see appendix \ref{sec:opt}.}. The estimation of $\beta_{jk}$ is further optimized following appendix \ref{sec:opt}.\\

\begin{figure}[!h]
  	\begin{center}
   	 \includegraphics[width=0.7\columnwidth]{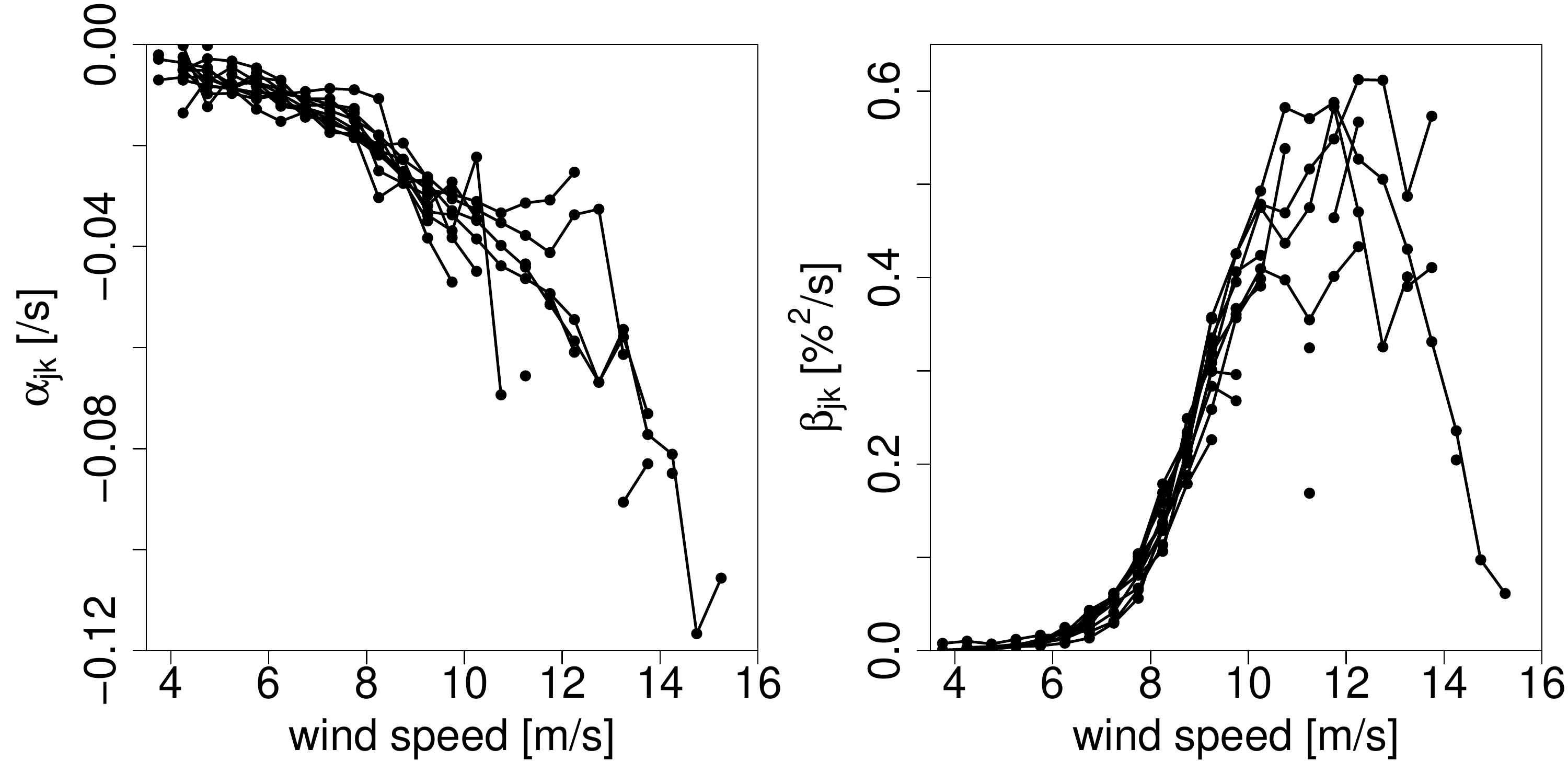}  
	\end{center}
	\caption{(left) slope of the drift coefficient $\alpha_{jk}$ and; (right) mean of the diffusion coefficient $\beta_{jk}$ for all direction bins $k$, as a function of the wind speed (bin j).}
  \label{fig:KMC_parameters}
\end{figure}

All the values of $\alpha_{jk}$ and $\beta_{jk}$ are presented in figure \ref{fig:KMC_parameters}. Both terms depend only weakly on the wind direction, especially at lower wind speeds where wake effects might be less pronounced. $\alpha_{jk}$ grows rather cubically, i.e. $\alpha_{jk}\propto u^3$. This indicates that the reaction time of the wind farm (the inverse of $\alpha$) drops cubically with $u$, stressing that the wind farm changes power values much faster at large wind speeds. the diffusion mean $\beta_{jk}\propto u^6$ to counteract the growing drift. At large wind speeds $u>13$m/s, the diffusion term drops because the pitch mechanism of the turbines regulates the power output that fluctuates very little. It is important to note that a larger diffusion term stresses the greater need for a stochastic model. At large wind speeds, the stochastic model becomes better-suited than a simpler deterministic model because the diffusion term is large. This illustrates the flexibility of a drift/diffusion model, where the drift quantifies the reaction time (inertia) of the system, and the diffusion/drift ratio controls the strength of the fluctuations.

The parametrization greatly reduces the number of model parameters. While the two estimated coefficients contain each $N_P \times N_u \times N_{\bar{\phi}}$ values, their two parametric forms only requires $N_u \times N_{\bar{\phi}}$ values each. This parametrization also gives more insight on the dynamics of the wind farm. The coefficients simplified with $\alpha_{jk}$ and $\beta_{jk}$ are used within the Langevin equation (\ref{eq:langevin}) to model the conversion process $(u(t),\bar{\phi}) \to P_2(t)$. Some results are presented in subsection \ref{sec:model_results} in comparison with other models.

In addition, a further simplification of the model could be achieved given the fact that $P_L\propto u^3$ for $P_L<P_r$. Then $\alpha_{jk}\simeq \alpha_{0} P_L$ and $\beta_{jk}\simeq \beta_{0} P_L^2$, where the parameters $\alpha_0$ and $\beta_0$ depend weakly on the wind direction. If this weak direction dependence is dropped, the Langevin model can be further simplified to the two-parameter form
\begin{eqnarray}
\frac{dP(t)}{dt}=\alpha_0 \, P_L\big(u(t)\big) \cdot \Big(P(t)-P_L(u(t))\Big)+\sqrt{\beta_0} \, P_L\big(u(t)\big) \cdot \Gamma(t) \, .
\label{eq:langevin_norm}
\end{eqnarray}
The two parameters are $\alpha_0\simeq -(6.48\pm 0.25)\times 10^{-4}\, \%^{-1}s^{-1}$ and $\beta_0\simeq (7.42\pm 0.21) \times 10^{-5}\, s^{-1}$ for the wind farm. The question of the dependence of these two parameters on the nature and size of the wind farm arises, and would be of interest for further studies. One should note that for better accuracy, $\beta_0$ should not stay constant but be reduced for $u>13$m/s to reproduce the effect of pitching. Results of the simplified model are not presented here.

\subsection{Pre-modeling of turbulent wind fluctuations}
\label{sec:model3}

Both models 1 and 2 replicate the conversion process $(u(t),\bar{\phi}) \to P(t)$. A practical application consists in predicting the power production when the wind condition is known. However, models 1 and 2 require a wind speed signal $u(t)$ sampled at around 1Hz and a ten-minute mean wind direction $\bar{\phi}$. Yet practical applications summarize the wind condition to ten-minute (or even one-hour) averages, and the 1Hz information is not known. In order to overcome such limitation, one must be able to model the wind dynamics at 1Hz realistically from the ten-minute (or even slower) wind data. Such a model is presented in this subsection.

We developed a stochastic model that generates high-frequency fluctuations of wind speed \footnote{The IEC norm \cite{IEC} defines wind speed fluctuations as a Gaussian field with spectral properties described by either a Kaimal or a von Karman spectrum.}. The model superimposes high-frequency fluctuations to the ten-minute average, following the description of atmospheric wind speed data from \cite{Boettcher2007,Morales2010a}. Simply put, the wind speed signal is decomposed as
\begin{eqnarray}
u(t) &=& \bar{u} + \sigma \cdot u'(t) \, \\
     &=& \bar{u} \, \big( \,1 + TI \cdot u'(t) \,\big) \, ,
\label{eq:udn}
\end{eqnarray}
where $\bar{u}$ and $\sigma$ are respectively the ten-minute mean value and standard deviation of the wind speed $u(t)$, and $TI$ is the turbulence intensity $TI=\sigma / \bar{u}$. In this description, the slow wind dynamics are described by $\bar{u}$ and $TI$ (that change every ten minutes), and the fast turbulent fluctuations are given by $u'(t)$ \footnote{The question whether wind dynamics can be easily separated into small-scale (turbulence) and large-scale (meteorology) effects is still an open question. The historical yet controversial hypothesis of a spectral gap \cite{vdHoven1957} implied such a separation. Some recent studies go beyond such simplification and present more complex descriptions such as e.g. multifractal cascade models.}. We observed that the fluctuations $u'(t)$ extracted from the measured wind speed $u(t)$ have a distribution reasonably close to the normal distribution \footnote{The normal distribution is commonly observed in natural phenomena, including the wind speed of some turbulent flows. Yet this oversimplified description of turbulence is deceived by a more complex multifractal scaling and small-scale intermittency. Here we consider the effective wind speed $u(t)$ averaged over the entire wind farm, whose fluctuations $u'(t)$ are close to a Gaussian monofractal process, see also Ref. \cite{Morales2010a}. We believe that a random Gaussian model is in this particular case acceptable.}. The autocorrelation function of $u'(t)$ is a rapidly decaying, exponential-like function
\begin{eqnarray}
R_{u'u'}(\tau)=\big\langle u'(t+\tau)u'(t)\big\rangle \simeq exp(-\gamma \tau).
\label{eq:udn_acf}
\end{eqnarray}
Such exponentially-correlated, Gaussian process is similar to the so-called {\it Ornstein-Uhlenbeck process} \cite{Risken1996} that is a special class of Langevin processes. The wind fluctuation signal $u'(t)$ can be modeled as an Ornstein-Uhlenbeck process following
\begin{eqnarray}
\frac{du'(t)}{dt}=-\gamma u'(t) + \sqrt{\gamma}\cdot \Gamma(t) \, ,
\label{eq:udn_OUP}
\end{eqnarray}
where $\Gamma(t)$ is a Langevin noise, that was already introduced in equation (\ref{eq:langevin}). Equation (\ref{eq:udn_OUP}) can be integrated numerically using the Euler scheme, see equation (\ref{eq:langevin_euler}).

In summary, we propose a simple model for the fluctuations $u'(t)$ of the wind speed signal $u(t)$ (representing the effective wind speed over the wind farm). One should be aware that turbulence is in general much more complex, as was shown e.g. by n-point statistics \cite{Stresing2010}. But here we show that this ansatz already works quite well for our particular application. While finer approaches can be developed, this simple method suffices here. $u'(t)$ is modeled by the stochastic equation (\ref{eq:udn_OUP}), where the unique parameter $\gamma$ is extracted from the measured wind speed following equation (\ref{eq:udn_acf}) \footnote{The decay parameter $\gamma$ should be estimated once for the given location of the wind farm as a measure of the local wind correlations.}. The signal $u'(t)$ is then used in combination with the measured ten-minute mean wind speed $\bar{u}$ and turbulence intensity $TI$ following equation (\ref{eq:udn}) to construct a synthetic signal of wind speed $u(t)$ at 1Hz. When this wind speed model is combined with the wind farm model presented in subsection \ref{sec:model2}, we can model the farm power output $P_3(t)$ at 1Hz knowing only the ten-minute mean wind speed, turbulence intensity and mean wind direction. Some results are presented in subsection \ref{sec:model_results} in comparison with other models.

\subsection{Statistical validation of the stochastic approach}
\label{sec:model_results}

We compare the power output signals modeled to the reference power output $P(t)$ measured. Three variations of our stochastic model are presented in subsections \ref{sec:model1}, \ref{sec:model2} and \ref{sec:model3}, yielding three power output signals $P_1(t)$, $P_2(t)$ and $P_3(t)$, see table \ref{table:models}. A fourth model is tested using the IEC power curve directly to convert at 1Hz the measured wind speed $u(t)$ into a power output $P_{IEC}(t)\equiv P_{IEC}(u(t))$. The five signals are compared in identical conditions over the total period of the measurement campaign (54 days spread over a total period of eight months, during which all turbines operate).

\begin{figure}[!h]
  	\begin{center}
   	 \includegraphics[width=0.6\columnwidth]{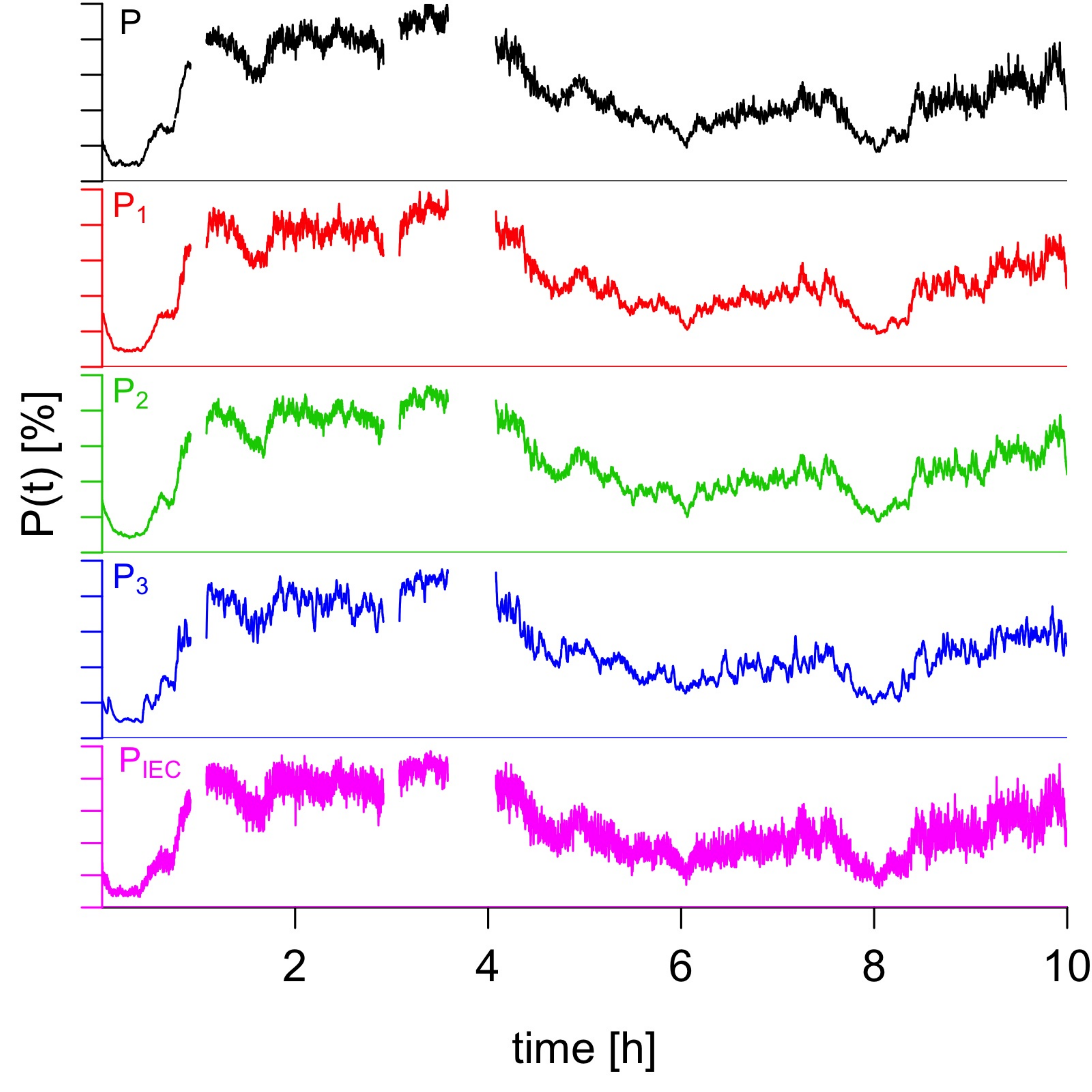}  
	\end{center}
	\caption{ Excerpts of power output signals sampled at 1 Hz over a continuous time period of 10h. The measured signal $P(t)$ is compared to the four modeled signals $P_1(t)$, $P_2(t)$, $P_3(t)$ and $P_{IEC}(t)$ (top to bottom). For each signal the vertical axis spans the range of power values from $0\%$ to $100\%$.}
  \label{fig:model_series}
\end{figure}

Excerpts of the five signals are presented in figure \ref{fig:model_series}. The time period shown was chosen because it covers a wide range of power values (the time series are displayed as a first illustration, whereas the statistical analysis covers the entire measurement period). The power signal measured $P(t)$ shows fluctuations at many scales and complex statistical features such as volatility clustering in the first five hours. Large power changes of about $50\%$ are observed within minutes. These dynamics measured are well reproduced by the first two stochastic models for $P_1(t)$ and $P_2(t)$. The third stochastic model for $P_3(t)$ also follows the slow trend with similar fluctuations, but shows some minor variations due to the fact that only ten-minute information of the wind was used (the wind fluctuations at 1 Hz were modeled, see subsection \ref{sec:model3}). The IEC method for $P_{IEC}(t)$ yields mixed results, as the slow trend is reproduced, but fast fluctuations are strongly over-represented. These results can be explained easily: the first two stochastic models reproduce the measured features because they exploit 1Hz wind data and they model the wind farm dynamics, including its inertia; the third stochastic model also models the farm dynamics, but only has access to ten-minute wind data, thus the fastest fluctuations are synthetic; finally, the IEC method does not model the fast farm dynamics but only the long-term behavior, as expected from using an average curve.\\

\begin{figure}[!h]
   \centering
   \begin{minipage}[t]{0.48\linewidth}
      \centering
      \includegraphics[width=0.8\linewidth]{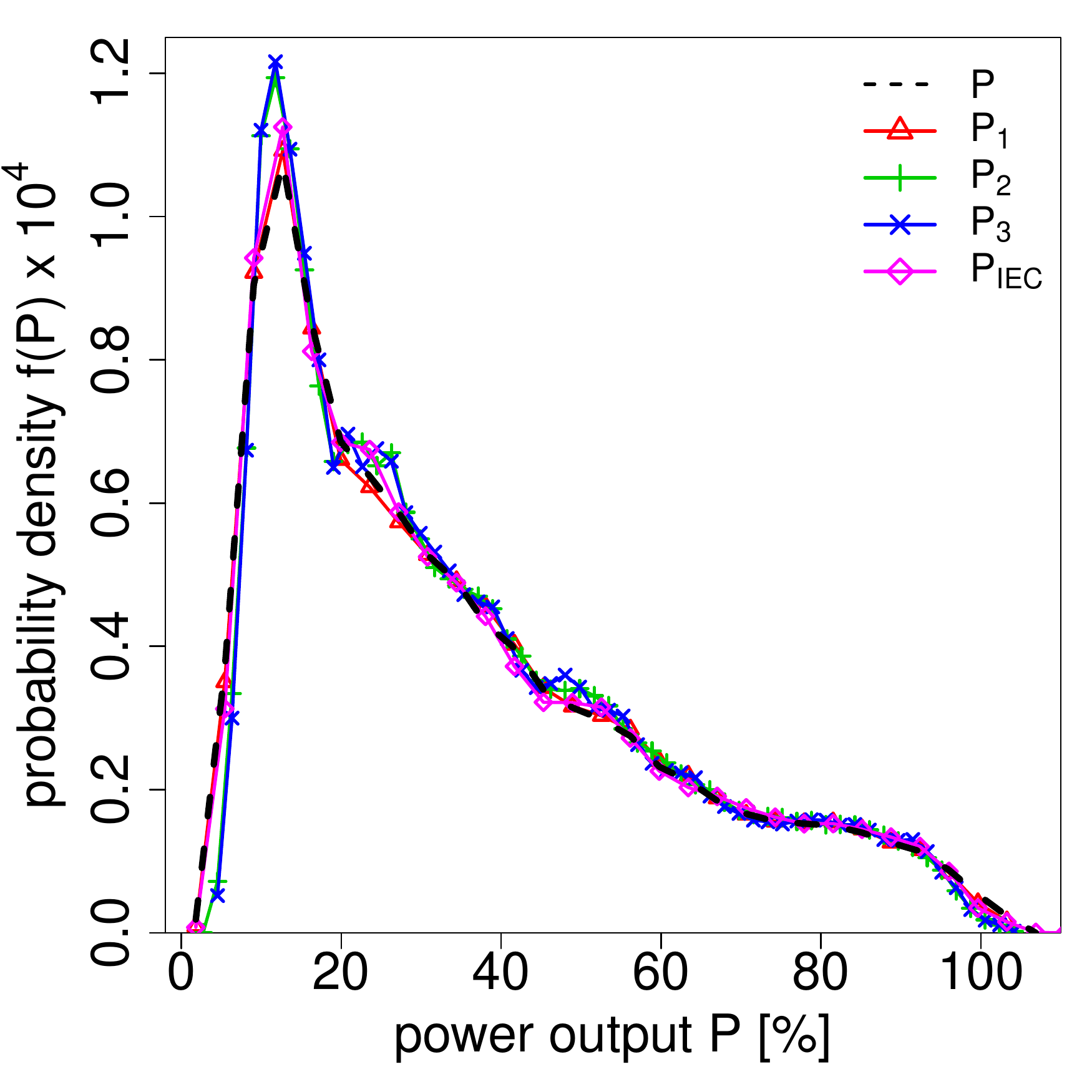}
      \caption{probability density functions (PDFs) $f(P)$ over the total measurement period. $f(P)$ was estimated using 50 bins.}
      \label{fig:model_hist}
   \end{minipage}%
   \hspace{0.5cm}%
   \begin{minipage}[t]{0.48\linewidth}
      \centering
      \includegraphics[width=0.8\linewidth]{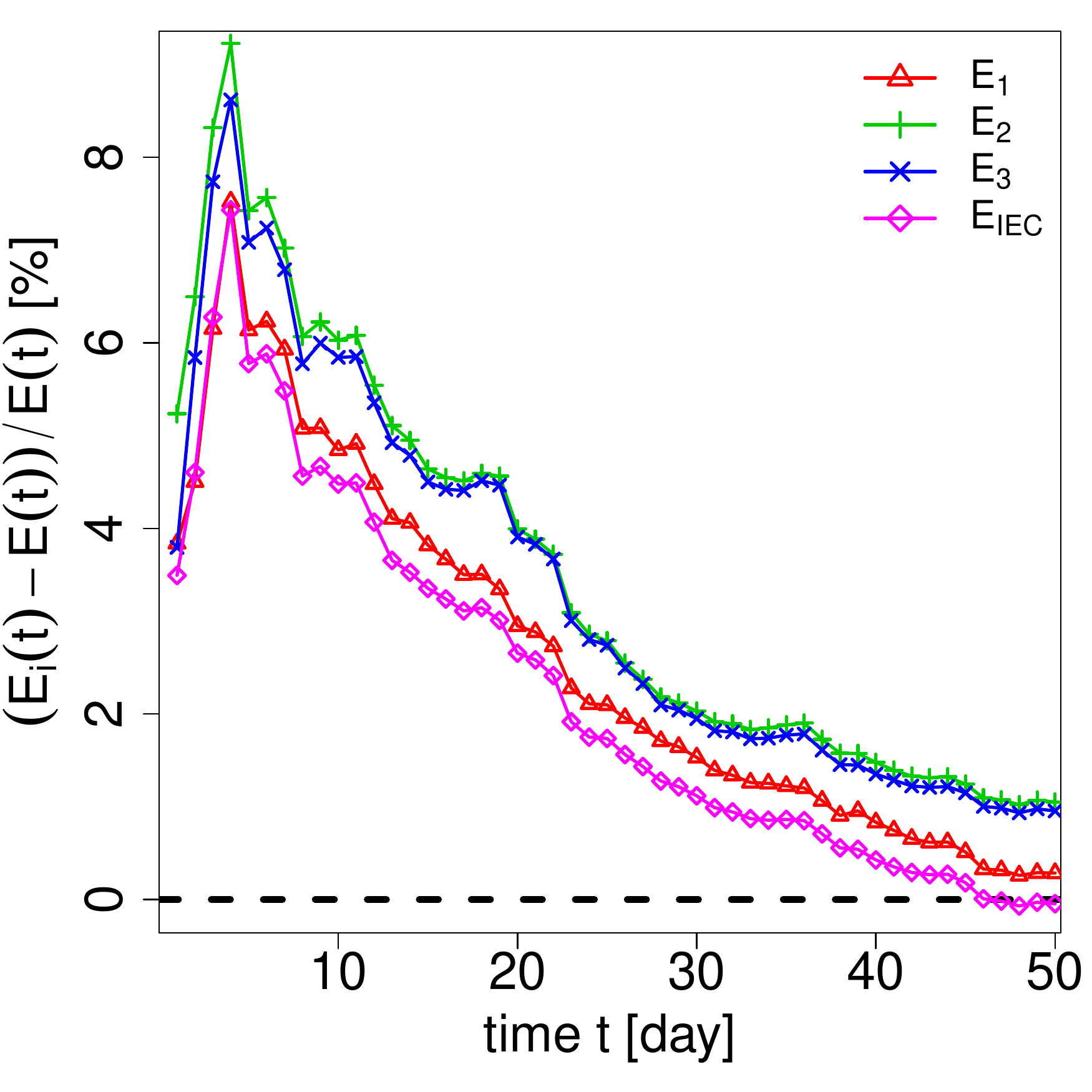}
      \caption{Relative difference of energy production $(E_i(t)-E(t))/E(t)$ for the various models $i=1,2,3,IEC$ in [$\%$].}
      \label{fig:model_energy}
   \end{minipage}
\end{figure}

The probability density functions (PDFs) $f(P)$ of the power output signals are presented in figure \ref{fig:model_hist}. Except for some minor deviations from the parametric results $P_2(t)$ and $P_3(t)$, all the models span the same values as the measurement with equal probabilities.

We present the cumulative energy produced as a function of time $E_i(t)=\int_0^t P_i(t')dt'$ in figure \ref{fig:model_energy} ($E(t)$ is calculated for the measured power $P(t)$). The four models produce some deviation of up to $ 9\%$ in the first 4 days of the simulation, then these deviations drop rapidly over longer integration times. The IEC model produces the best results all along for long-term energy production, and matches mostly perfectly the measurement over the total simulation time of 54 days, with a deviation of only $0.02\%$. It is to be expected that the IEC power curve, being an average curve, can reproduce the total energy (i.e. mean power) of the data set from which it is estimated. It is followed closely by model 1 (estimated), that only deviates by $0.3\%$ of the total energy production over the total simulation time. Models 2 and 3 show a slight deviation of $0.8\%$ and $0.7\%$ from the measurement over the simulation time. While the IEC power curve remains optimal for long-term energy production, minor deviations are found using the estimated model 1. Larger deviations are introduced by the parametrization in models 2 and 3. The stochastic models show satisfactory results, considering that an integration over about 5 million stochastic samples deviates from the measurement at most by $1\%$. This is an important validation that the stochastic models follow the measurement over time without diverging (as tested tens of times for the three stochastic variations). This is due to the nature of the model, and particularly of the drift field that constantly drives the dynamics towards the power curve. Properly reproducing slow dynamics is a necessary (but not sufficient) condition for a model of the high-frequency dynamics. The integral power is a first-oder quantity of the power conversion process. Next we aim to present a deeper statistical analysis of the data. In particular we focus on two-time quantities, namely the power spectrum, the autocorrelation function as well as the probability distribution of power increments and its corresponding structure functions, for further details on the analysis scheme see Ref. \cite{Morales2010a}. \\

\begin{figure}[!h]
   \centering
   \begin{minipage}[t]{0.48\linewidth}
      \centering
      \includegraphics[width=0.8\linewidth]{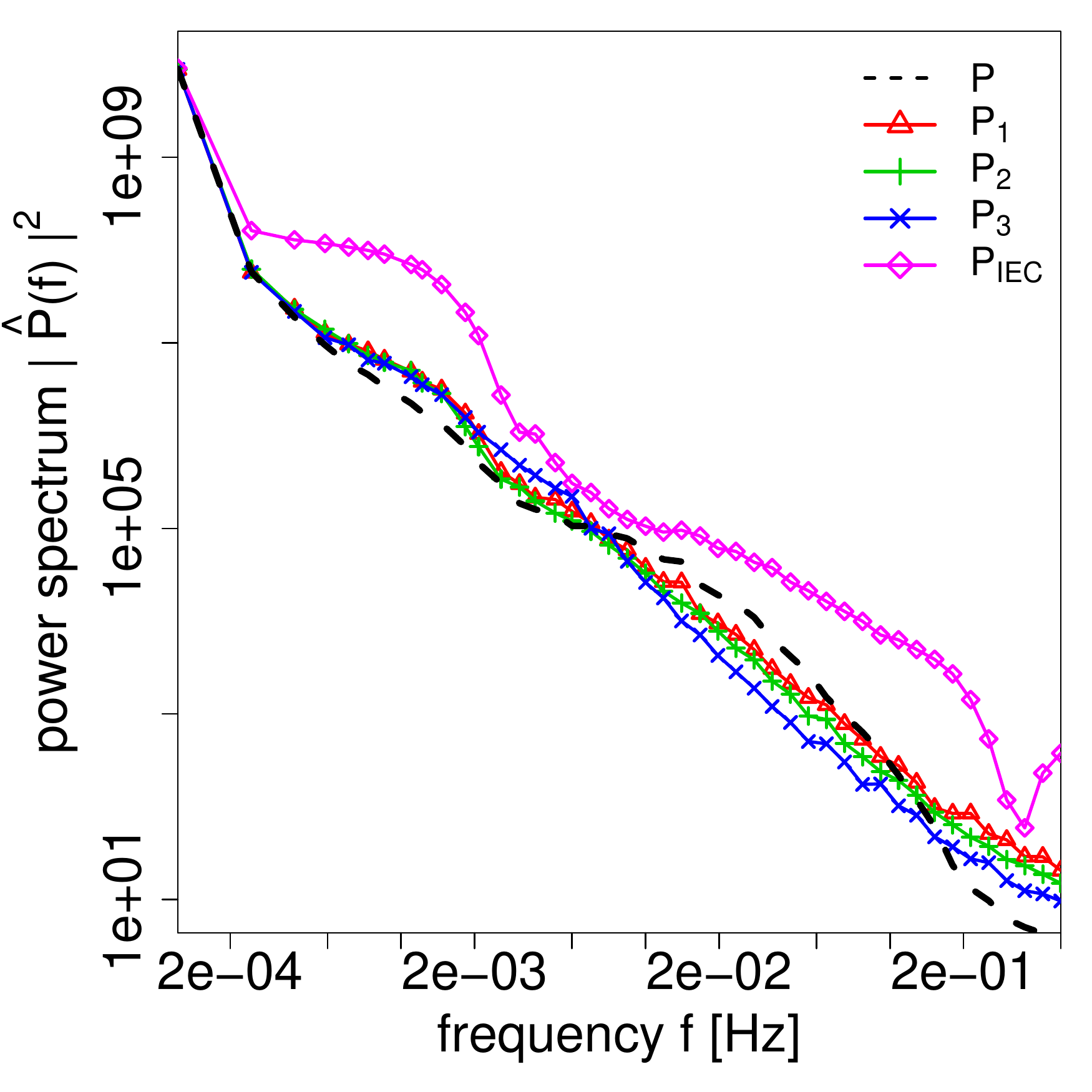}
      \caption{power spectral density $|\hat{P}(f)|^2$ for the five power signals, displayed in log-log scale. The frequency range is $f \in [2^{-13},2^{-1}]$ Hz.}
      \label{fig:model_spec}-
   \end{minipage}%
   \hspace{0.5cm}%
   \begin{minipage}[t]{0.48\linewidth}
      \centering
      \includegraphics[width=0.8\linewidth]{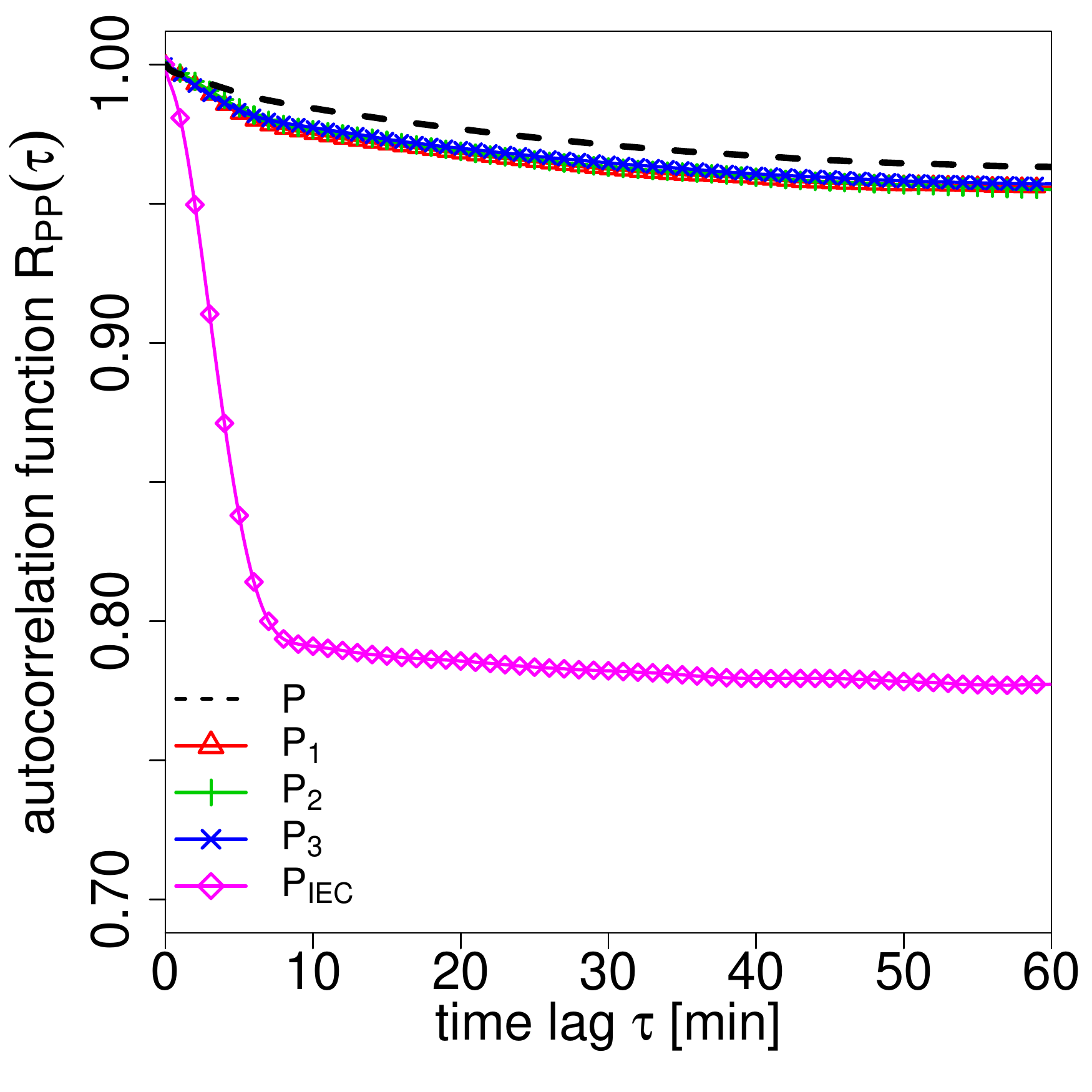}
      \caption{autocorrelation function $R_{PP}(\tau)$ for the five power signals, displayed in linear-log scale as a function of the time lag $\tau$.}
      \label{fig:model_acf}
   \end{minipage}
\end{figure}

The power spectral density commonly referred to as {\it spectrum} $|\hat{P}(f)|^2$ \footnote{The Fourier transform $\hat{P}(f)$ was calculated using the {\it Fastest Fourier Transform in the West} algorithm. The spectrum was then estimated using Bartlett's method to obtain a denoised spectrum \cite{Press2007}.} is presented in figure \ref{fig:model_spec} for the power signals. Each spectral value $|\hat{P}(f)|^2$ indicates the energy of the corresponding wave of frequency $f$. The spectra can be described reasonably well by a power law $f^{-\beta}$ (linear behavior in log-log scale). The three stochastic models have similar spectra, very close to a power law. They are close to that of the measurement, although the measured spectrum is more complex, with different scaling behaviors in various frequency bands. The IEC power curve model deviates from the measurement, having spectral values up to two orders of magnitude too high; it over-represents the intensity of the fluctuations for $f>2.10^{-4}$Hz, corresponding to fast fluctuations up to roughly 80 minutes. Although not shown here, all spectra match at lower frequencies $f<2.10^{-4}$Hz, indicating that they all follow the trend of the wind dynamics at 80 minutes and longer time scales.

The spectrum is related to the autocorrelation function $R_{PP}(\tau)$ \footnote{The autocorrelation function of a stationary signal $x(t)$ is given by $R_{xx}(\tau)=\frac{\langle (x(t+\tau)-\bar{x})(x(t)-\bar{x}) \rangle}{\sigma_x \,^2}$, where $\sigma_x \,^2$ is the variance of $x(t)$. $R_{xx}(\tau)$ is the inverse Fourier transform of the spectrum $|\hat{x}(f)|^2$ of $x(t)$, following Wiener-Khinchin theorem. Because the power signals we study are non-stationary, we calculated $R_{PP}(\tau)$ over time windows of $8192$s.}. The autocorrelation functions of the power signals are presented in figure \ref{fig:model_acf}. The power output measured has long time power-law-like correlations. As an example, a power output signal remains correlated to $97\%$ with itself when shifted by one hour. The three stochastic models match the measurement very well, indicating that they reproduce the time correlations of the measured power. They display power-law correlations close to that of the measurement, as already observed for the power spectral densities.
In contrary to the good results of the stochastic models, the IEC power curve model deviates from the measurement and generates a signal that is less correlated. This deviation can be related to results of the power spectral density, where the IEC model was shown to deviate significantly from all the other signals. Particularly, the large drop of correlations at $\tau<10$min corresponds mostly to spectral deviations at $f<2.10^{-3}$Hz, as the low-frequency Fourier modes control the coarse-grained dynamics.

\begin{figure}[!h]
  	\begin{center}
   	 \includegraphics[width=0.7\columnwidth]{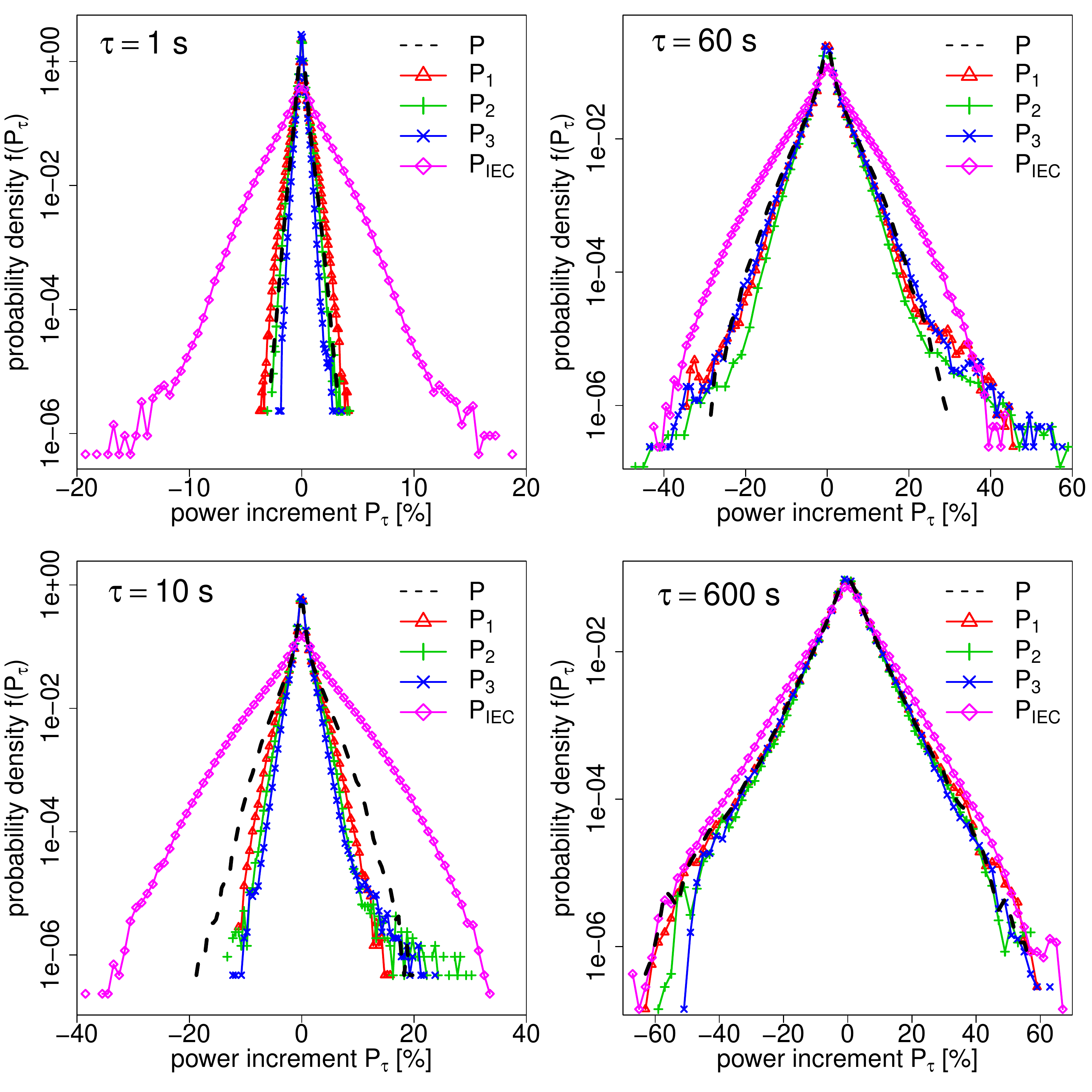}  
	\end{center}
	\caption{increment PDFs  $f(P_\tau(t))$ at various scales $\tau$ for the five power signals, displayed in linear-log scale.}
  \label{fig:model_incpdf}
\end{figure}

Our final statistical test consists in comparing the PDFs of the power increment $P_\tau(t)=P(t+\tau)-P(t)$. The power increment $P_\tau(t)$ is the change (increment) of power output $P(t)$ when waiting a time $\tau$. The PDF $f(P_\tau(t))$ indicates with what probability the power output can change over a time scale $\tau$ \footnote{The second-order 2-point moment $<P_\tau(t)^2>$ corresponds to the autocorrelation and the power spectrum. The PDFs $f(P_\tau(t))$ incorporate information about all the higher-order 2-point statistics $< P_\tau(t)^n>=\int P_\tau(t)^n f(P_\tau) dP_\tau$. This allows for a thorougher validation of our models.}. The PDFs of the power increments are presented in figure \ref{fig:model_incpdf} for four time scales. The power output measured has exponentially-distributed increments (linearly decaying tails in linear-log scale). This is typical for non-Gaussian, intermittent processes that have some quiet time periods alternating with some periods of strong fluctuations, as observed for the time series in figure \ref{fig:model_series}. This notion relates to the intermittent nature of the wind \cite{Morales2010a} that is converted into power intermittency by the wind farm \cite{Milan2013a}.
At the shortest time scale available $\tau=1$ s, the stochastic models reproduce the increment PDF of the measurement mostly well, with a slight over-representation of large power increments for model 1. Model 2 and 3 match the measurement. The IEC procedure generates much too large power fluctuations at this time scale, with increments up to $\pm 20\%$ within one second, when the measurement can change by at most $\pm 5\%$.
At the time scale $\tau=10$ s, the stochastic models under-represent large power increments and the IEC procedure still largely over-represents them. The measured power output can change by $\pm 20\%$ within ten seconds.
At the next time scale $\tau=60$ s, the stochastic models match the measurement very well, except for slightly too many extreme events at $P_\tau>30\%$, and the IEC procedure comes closer to the measurement. Note that for these strong fluctuations the probabilities are obtained from 10 and less events per bin, thus error bars larger than $30\%$ are expected. The measured power output can change by $\pm 30\%$ within one minute.
At the time scale $\tau =600$ s, the four models match the measurement very well, except for the IEC procedure that still slightly over-represents extreme events. The measured power output can change by $\pm60\%$ within ten minutes. We want to stress that the probabilities also show that more extreme events are expected when longer data sets are considered.

Note, a power fluctuation of $5\%$ corresponds to a change of more than $1$MW here. Fluctuations of several MW may effect the local grid stability essentially. The non-Gaussianity of the increment PDFs tells us that even more extreme values become likely over longer time periods. This is important for grid stability and puts additional constraints on the design of back up methods such as energy storage or overload security. Power gustiness is an important feature of wind power production. Rapid power changes affect grid stability, especially on networks that rely strongly on wind energy. When considering the entire dataset, we observe power changes of up to $60\%$ within 3 minutes, and up to $85\%$ within 10 minutes. This stresses the importance of power intermittency on grid stability, and how necessary it is to properly model it.

We promote the three stochastic models that generate some power gusts that have a similar amplitude to that of the power gusts measured. These gusts also occur with similar probabilities. Thus our model time series are clearly better than the time series obtained from the IEC procedure, which strongly exaggerates fast power gusts in time scales of seconds. For time scales of ten minutes and more both methods become similar. This confirms that the IEC procedure is valid for slow dynamics, and fluctuates too strongly for fast dynamics within one hour.

\begin{figure}[!h]
   \centering
   \begin{minipage}[t]{0.48\linewidth}
      \centering
      \includegraphics[width=0.8\linewidth]{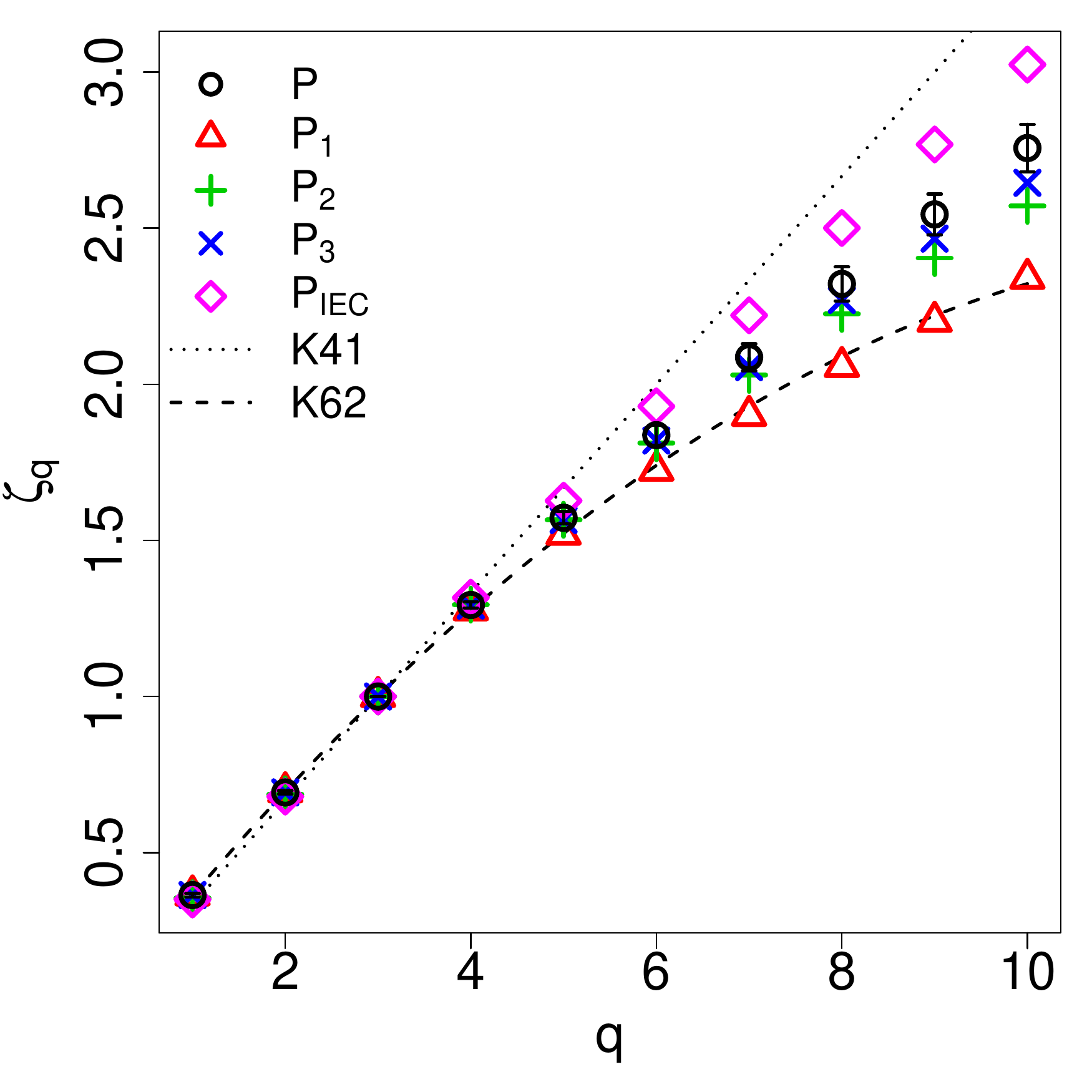}
      \caption{scaling exponents $\zeta_q$ as a function of the order $q$ for the five power signals, calculated for scales in the range $\tau \in \{256,16384\}$ s. K41 (dotted line) and K62 (dashed line) are given for reference.}
      \label{fig:model_zeta}
   \end{minipage}%
   \hspace{0.5cm}%
   \begin{minipage}[t]{0.48\linewidth}
      \centering
      \includegraphics[width=0.8\linewidth]{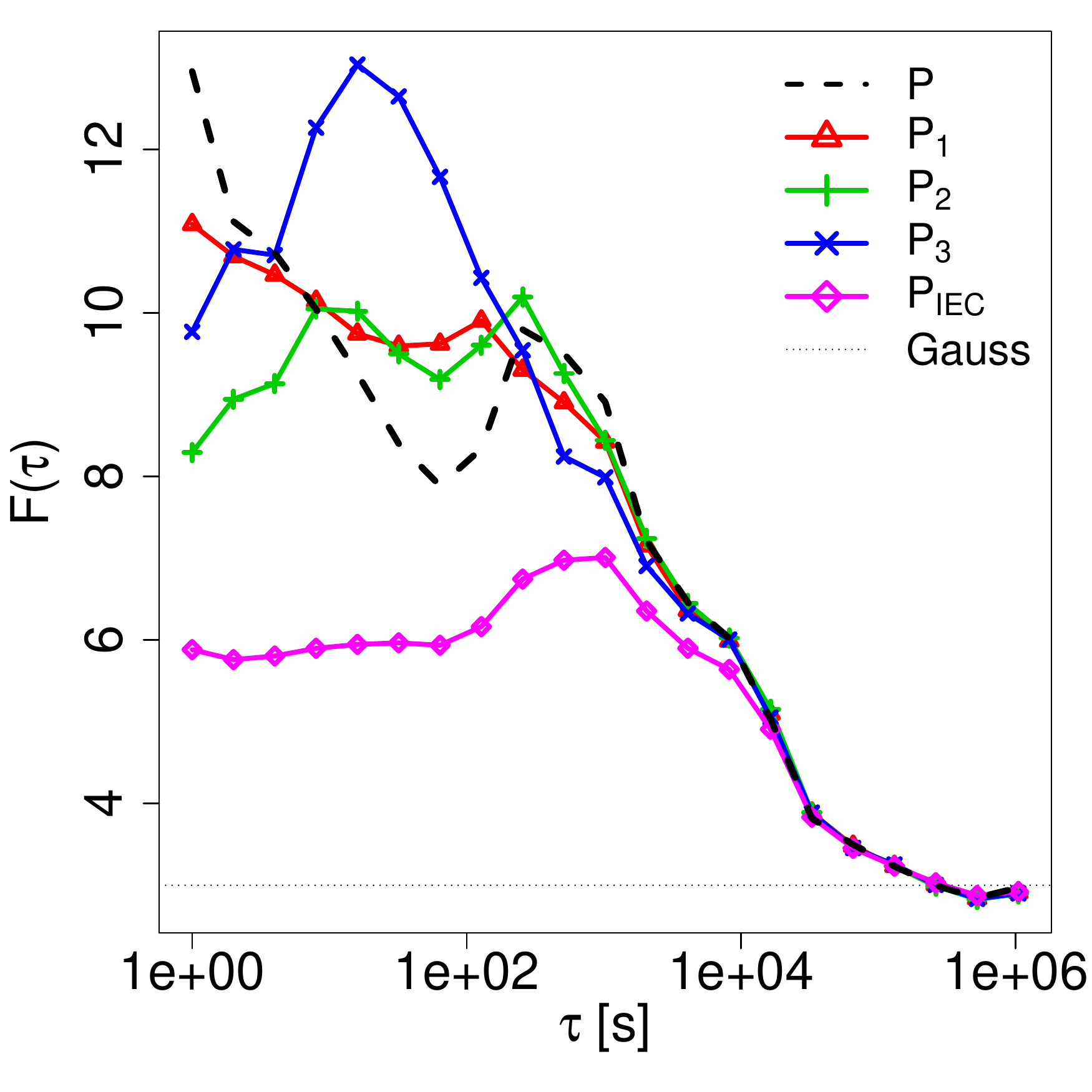}
      \caption{flatness $F(\tau)$ for the five power signals, displayed in log-linear scale as a function of the time lag $\tau$. $F=3$ indicates a Gaussian process (horizontal dotted line).}
      \label{fig:model_flatness}
   \end{minipage}
\end{figure}

Gustiness and intermittency are related to the scaling of the increments $P_\tau$. We define the structure functions following \cite{Arneodo1996}
\begin{eqnarray}
S_q(\tau)=\langle |P_\tau|^q \rangle \sim \tau^{\zeta_q}.
\label{eq:strucfunc}
\end{eqnarray}
The power law scaling $S_q(\tau) \sim \tau^{\zeta_q}$ is observed for all the power output signals, indicating their fractal nature. Additionally, the scaling exponent $\zeta_q$ indicates the nature of the fractality \footnote{More precisely, we use the {\it extended self-similarity} method from Ref. \cite{Benzi1993} to remove a slightly anomalous scaling observed in the structure functions. This method extracts the scaling exponents from the relation $S_q(\tau) \sim S_3(\tau)^{\zeta_q}$, forcing $\zeta_3=1$.}. A linear dependence $\zeta_q \propto q$ indicates a monofractal signal with self-similar features along scales, implying that the form of the PDFs $f(P_\tau)$ would not change. In this case only the $\tau$-dependency of the variance would be needed to characterize the statistics. If $\zeta_q$ is instead a nonlinear, concave function of $q$, the process is multifractal with an intermittent, walk-like behavior at small scales and a noise-like behavior at large scales. Consequently we need to characterize $f(P_\tau)$ for all $\tau$. The scaling exponents are presented in figure \ref{fig:model_zeta}, compared to Kolmogorov's 1941 monofractal and 1962 multifractal models for ideal local isotropic turbulence.
The measured power lies in between the two models, being intermittent but less than K62. Models 2 and 3 reproduce very well the multifractality of the measurement. Model 1 is too multifractal, probably owing to the (still too) quadratic form of the diffusion coefficient estimated. We observe that model 3 (that uses Gaussian wind fluctuations) is multifractal, showing that power intermittency for the wind farm comes from the conversion process, rather than from the intermittency in the wind. The IEC model is less multifractal, as the small-scale intermittency introduced by the inertia of the wind turbine rotors is not reproduced.

The flatness (or kurtosis) $F(\tau)=\frac{S_4(\tau)}{(S_2(\tau))^2}$ \cite{Press2007} quantifies directly the non-Gaussian character of the distribution $f(P_\tau(t))$, see figure \ref{fig:model_flatness}. We measure a high flatness up to a value of $12$ at short time scales in seconds confirming the strong intermittency at short time scales. This value is reduced by a factor two within two hours, and reaches the Gaussian limit $F=3$ for $\tau>3$ days \footnote{It is interesting to note that the correlation length of some atmospheric wind datasets was found to be around $4-5$ days in \cite{Kavasseri2004,Muzy2010a}, matching our observation for the wind farm output. The autocorrelation function shows $R_{PP}(\tau=4$days$) \simeq 0.25$ and $R_{PP}(\tau>15$days$) \simeq 0$.} \footnote{A local peak is observed for the flatness of the measured data for $\tau \sim 10$ minutes. This time scale corresponds to the characteristic size of the farm of about $3$ km. It could be due to the fact that the farm converts atmospheric structures at this scale without filtering them, yielding a local increase in intermittency. This feature is not reproduced by any of the models.}. Model 1 matches the measurement moderately well for $\tau<100$ s, and models 2 and 3 deviate more at those scales. At larger scales $\tau>200$ s, the three stochastic models match the measurement very well. The IEC model is not intermittent enough for $\tau<3$ hours, then matches the measurement at larger scales.

\section{Performance monitoring}
\label{sec:monitor}
After characterizing and modeling the power output based on the wind speed, the stochastic approach has a second important application for the overall monitoring of the wind park performance based on 1Hz data. Performance monitoring consists in tracking the power performance of the wind installation. A monitoring procedure verifies in real time whether the installation is running in its optimal configuration, or if maintenance is needed to fix a potential anomaly. Many components of e.g. aerodynamic, mechanical or electrical nature interact in complex ways in a wind turbine. A detailed monitoring of each single component yields a huge amount of data to analyze in real time. This {\it microscopic} monitoring can be advantageously simplified to a {\it macroscopic} procedure, where a set of global estimates describes the overall system. This reduction of the many variables into a set of global estimates is even more meaningful for a large array of wind turbines such as a wind farm.

The reduced amount of information must be balanced by insightful estimates that properly extract the essential information. These estimates must be sensitive enough to detect anomalies, yet robust enough to cope with the expected fluctuations. The wind speed/power output signals are commonly displayed together as a power curve. The IEC power curve (see section \ref{sec:dyn10min}) gives a rapid overview of the overall performance of the conversion process $u \to P$ for a single wind turbine. The extensive averaging applied greatly reduces the amount of data, but seriously lessens the sensitivity to fine changes. Based on our analysis above, the IEC power curve is expected not to be an optimal estimate of the dynamical performance. Instead, we promote the drift coefficient $D^{(1)}(P,u;\phi)$ (see section \ref{sec:model1}) as a flexible and accurate alternative.

Our goal here is to test the sensitivity of both the IEC and the stochastic estimates, see resp. subsections \ref{sec:monitor_iec} and \ref{sec:monitor_drift}. First we want to find out how long an anomaly must last, so that it is surely detected. In particular, the progressive modification of the estimates is studied in relation to the anomaly duration for both methods. As a second aspect, the impact of continuous (uninterrupted) anomalies as well as intermittent anomalies (in the sense that they only last during short time periods, here of 1 min) is analyzed.

A dataset is selected within the wind sector $k=8$ ($\bar{\phi}\simeq 225^\circ$) spanning a discontinuous time period of 14 days, when all 12 turbines function properly. For each anomaly, we artificially shut down one or two wind turbines (set their power output $P_i(t)=0$) during a given duration and recalculate the cumulative farm power. We developed around 50 anomaly scenarios and tested their impact on the IEC and stochastic estimates. Four of these scenarios are selected here to illustrate the two central aspects: how long must the anomaly last; and how robust are the IEC and stochastic methods to anomalies occurring intermittently during short time periods? A proper definition of the four scenarios is presented in table \ref{table:scenarios}.
\begin{table}[!h]
  \centering
   \begin{tabular}{ c | c | c | c | c | c }
   \hline
    anomaly & description & shut down & anomaly duration & normal operation & anomaly type \\ \hline \hline
    (a) & 1shutdown\_12h 	& 1 turbine & 12h ($3.6\%$) & 13d12h ($96.4\%$) & continuous \\ \hline
    (b) & 1shutdown\_4d 	& 1 turbine & 4d ($28.6\%$) & 10d ($71.4\%$) 	& continuous \\ \hline
    (c) & 1shutdown\_4d\_int& 1 turbine & 4d ($28.6\%$) & 10d ($71.4\%$) 	& intermittent \\ \hline
    (d) & 2shutdown\_7d		& 2 turbines& 7d ($50\%$) 	& 7d ($50\%$) 		& continuous \\ \hline
  \end{tabular}
   \caption{overview of the four anomaly scenarios implemented over a 14-day time period. For each scenario (a)-(d), a description is given, as well as how many of the twelve turbines are shut down in the wind farm. The duration of the anomaly and of the normal operation are given in time (and in percent). Finally the anomaly type is given, whether continuous in time (at the end of the 14-day period) or intermittent (5760 one-minute-long anomalies are spread over the total time period, representing a total anomaly time of 4 days).}
 \label{table:scenarios}
\end{table}

\begin{figure}[!h]
  	\begin{center}
   	 \includegraphics[width=\columnwidth]{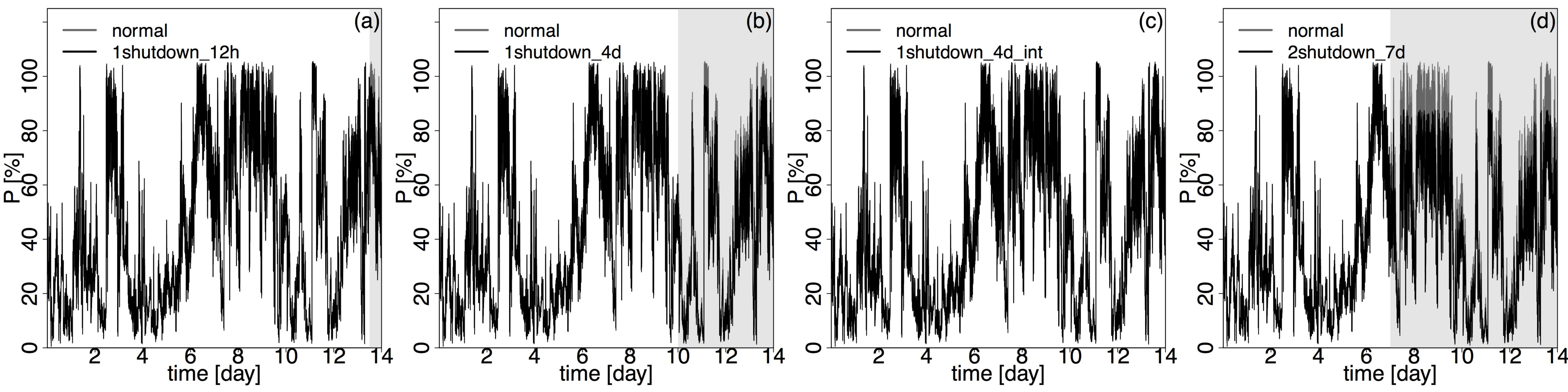}  
	\end{center}
	\caption{Wind farm power output $P(t)$ for the wind sector $k=8$ ($\bar{\phi}\simeq 225^\circ$) for scenarios (a-d) (left to right). In each anomaly scenario, the power output (black line) is compared to the power output measured (gray line in background). The period of the anomaly is marked by the gray area (except for (c) where the anomaly happens intermittently all over the 14-day period).}
  \label{fig:compare_monitoring_signal}
\end{figure}

We monitor the cumulative farm power output for each scenario, see figure \ref{fig:compare_monitoring_signal} and compare it to the measured (normal) farm output. It is important to note that the wind/power conditions constantly change over the 14-day period. The four anomalies cover different time periods, such that each scenario affects the power performance at different locations in the $\{u;P\}$ space. It makes it difficult to compare different scenarios from one another. Instead, we separately compare each scenario from the measured normal case and a {\it full anomaly} case where the shut down (of one or two turbines) occurs during the entire 14-day period.

\subsection{Monitoring procedure based on the IEC power curve}
\label{sec:monitor_iec}

\begin{figure}[!h]
  	\begin{center}
   	 \includegraphics[width=\columnwidth]{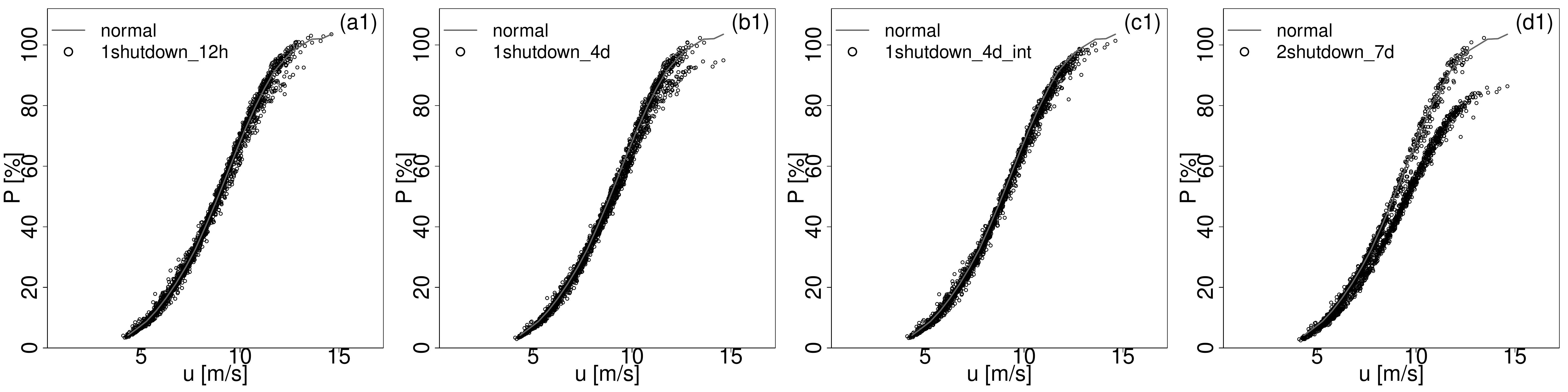}  
  	 \includegraphics[width=\columnwidth]{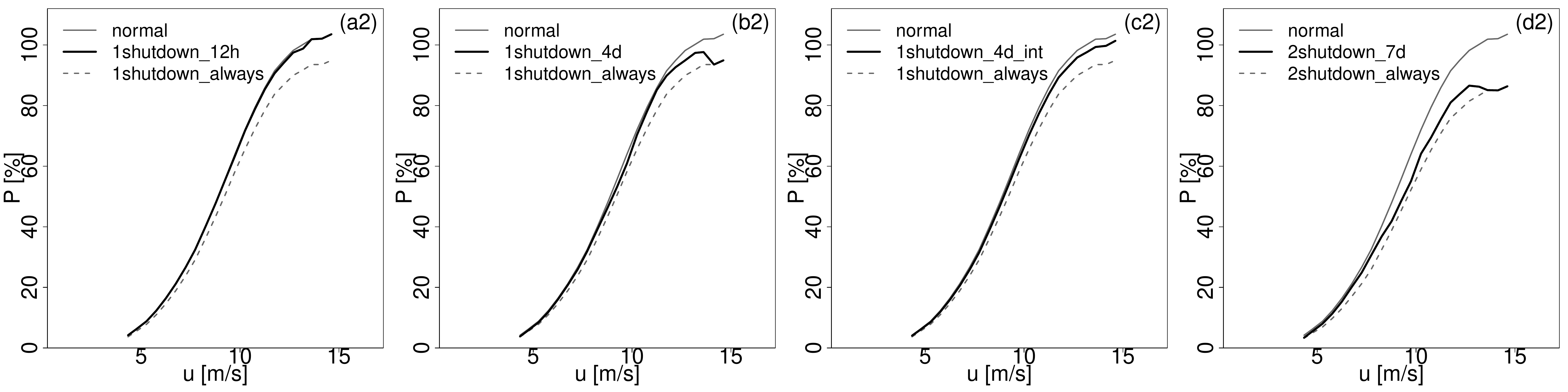}  
 	\end{center}
	\caption{(upper) ten-minute means of farm power $\bar{P}$ versus wind speed $\bar{u}$ (dots) and; (lower) IEC power curve $P_{IEC}(u_j,\bar{\phi}_k)$ (bold black line) for the wind sector $k=8$ ($\bar{\phi}\simeq 225^\circ$) for scenarios (a-d) (left to right). The IEC power curves are also given for the measured normal case (full gray line) and for the full anomaly case (dashed gray line).}
  \label{fig:compare_monitoring_iec}
\end{figure}

We compare in figure \ref{fig:compare_monitoring_iec} the impact of each anomaly on the ten-minute means of power output $\bar{P}$ and on the corresponding IEC power curve $P_{IEC}(u_j,\bar{\phi}_k)$.
We observe that for scenario (a), most ten-minute means lie around the power curve of the normal case, and the resulting IEC power curve in figure \ref{fig:compare_monitoring_iec}(a2) does not deviate noticeably from the normal case. The anomaly only lasts $3.6\%$ of the total time, which is not enough for the IEC procedure to detect it.
For scenario (b), enough ten-minute means deviate from the normal power curve, such that the IEC power curve is changed significantly by a step-like structure at wind speeds higher than $10$ m/s. We found that this is the minimal anomaly duration, i.e. $28.6\%$ of the time for the IEC method to detect an anomaly.
In scenario (c), the anomaly also lasts 4 days but intermittently. In this case the ten-minute means deviate only slightly from the normal case, and no outliers are observed. The resulting IEC power curve does deviate slightly from the normal case, but less so than for the equally long but uninterrupted anomaly in scenario (b). This is expected from a ten-minute averaging that mixes 9 minutes of normal operation with 1 minute of anomaly. This is a limitation of the IEC procedure, that cannot separate scenarios alternating faster than ten minutes.
Finally for scenario (d) two turbines are shut down during $50\%$ of the time. Half of the ten-minute samples are outliers that deviate clearly from the normal case. The IEC power curve lies in between the two states, as each state takes place half of the time. The anomaly is clearly detected.

\subsection{Monitoring procedure based on the drift coefficient}
\label{sec:monitor_drift}

The drift coefficient $D^{(1)}(P,u_j,\bar{\phi}_k)$ presented in section \ref{sec:model1} quantifies the conversion dynamics of the system to a first order approximation. It is a good candidate to detect fine changes in the farm dynamics. Additionally, its potential $\Psi(P,u_j,\bar{\phi}_k)$ gives a more intuitive representation of the attractors of the system, that define the Langevin power curve $P_L(u_j,\bar{\phi}_k)$.

\begin{figure}[!h]
  	\begin{center}
   	 \includegraphics[width=\columnwidth]{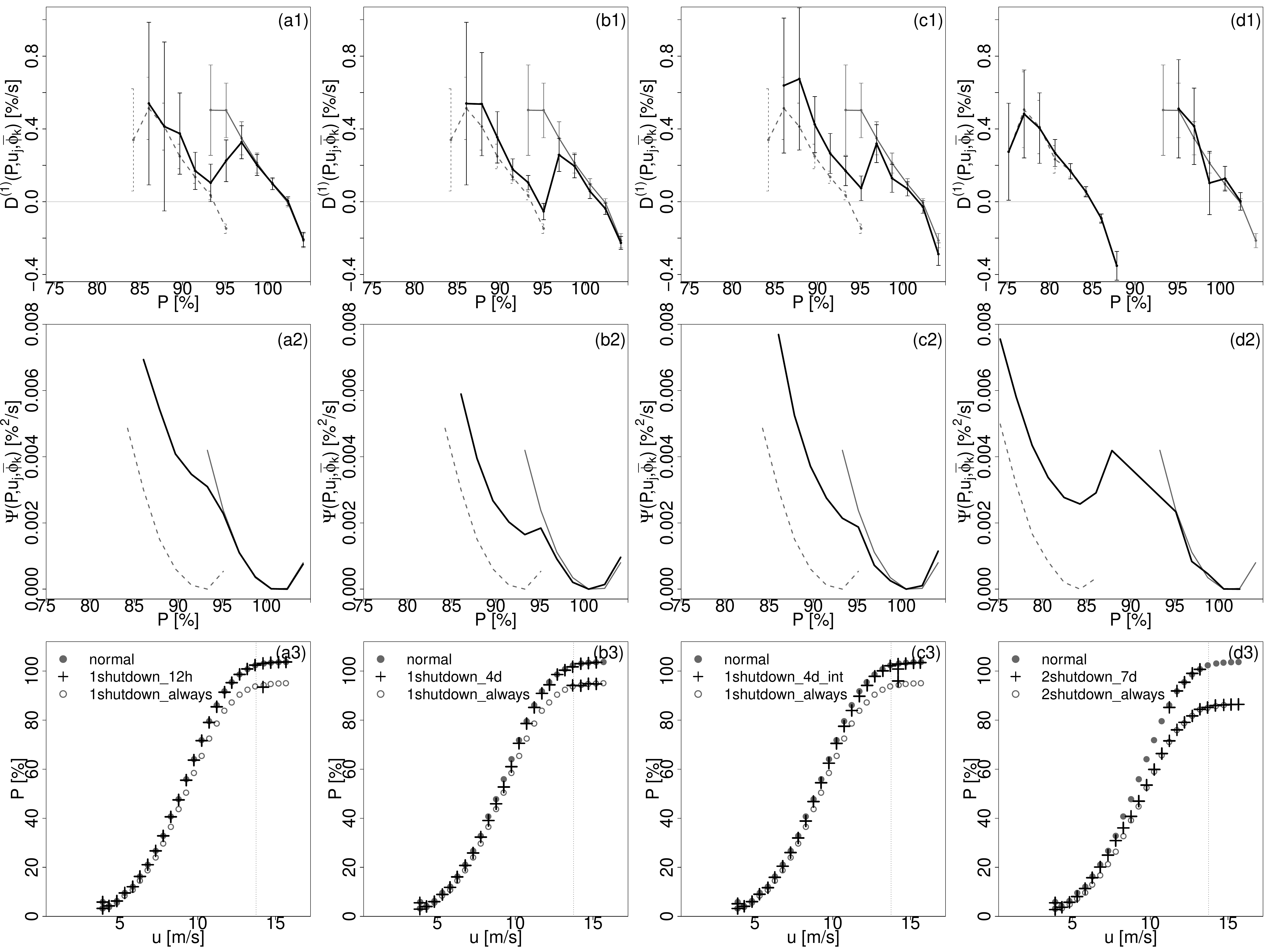}  
	\end{center}
	\caption{(upper) drift coefficient $D^{(1)}(P,u_j,\bar{\phi}_k)$ and; (middle) drift potential $\Psi(P,u_j,\bar{\phi}_k)$ for the wind speed bin $j=28$ ($u\sim 13.75$m/s); (bottom) Langevin power curve $P_L(u_j,\bar{\phi}_k)$ for the wind sector $k=8$ ($\bar{\phi}\simeq 225^\circ$) for scenarios (a-d) (left to right). In each plot, the anomaly case (bold black line / black crosses) is compared to the normal case (gray line / full gray dots) and to the full anomaly case (dashed gray line / open gray dots). The drift potential is arbitrarily shifted upwards for comparison (minimum set to zero). In the (lower) figures, the dotted vertical line indicates the wind speed bin $j=28$.}
  \label{fig:compare_monitoring_drift}
\end{figure}

We compare the impact of each scenario on the drift coefficient, potential and Langevin power curve in figure \ref{fig:compare_monitoring_drift}. As seen in figure \ref{fig:drift_potential}, the measured normal case (full gray line) corresponds to a linear drift towards the Langevin power value. The extreme case of a full anomaly (dashed gray line) also displays a fairly linear drift towards a new, reduced power value. Shutting down some turbines reduces the power production, and the farm naturally drifts towards a lower power value.
In the four scenarios, the normal and anomaly situations coexist, and the drift coefficient systematically records these two states. The anomaly duration acts as a weighting factor that determines how much the drift is distorted close to the reduced, anomalous power value.
For scenario (a), $3.6\%$ anomaly time is sufficient to modify the drift coefficient and potential noticeably at power values $P<95\%$.
We observe for scenario (b) that a longer anomaly time of $28.6\%$ brings new fixed points at $P\simeq 95\%$ in the rated power region. The drift coefficient displays multiple stable fixed points, and the potential has several local minima.
The intermittent anomaly in scenario (c) is also detected in the drift coefficient and potential. Less fixed points are created in the Langevin power curve, due to the fact that the anomaly scatters over the entire state space $\{u;P\}$. It is not as localized in the state space as the continuous anomaly in scenario (b), so that it affects a larger region of the state space, but to a lesser extent.
Finally for scenario (d), both dynamical states are recorded by the drift coefficient that clearly detects the anomaly superposed to the normal operation.

\begin{figure}[!h]
  	\begin{center}
   	 \includegraphics[width=0.5\columnwidth]{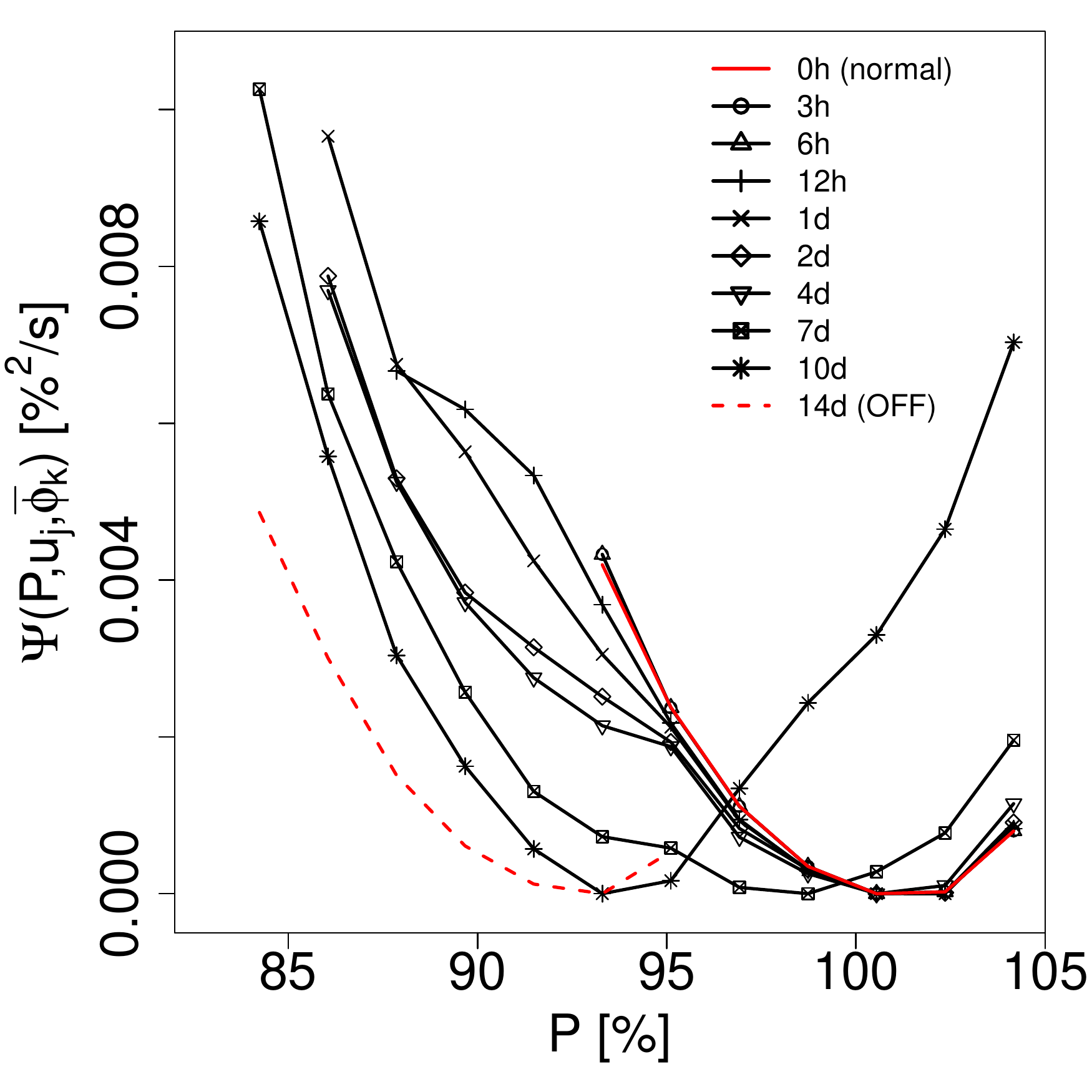}  
	\end{center}
	\caption{Drift potential $\Psi(P,u_j,\bar{\phi}_k)$ for the wind speed bin $j=28$ ($u\sim 13.75$m/s) and wind sector $k=8$ ($\bar{\phi}\simeq 225^\circ$) for an intermittent anomaly of varying duration (see legend). The drift potentials are given for the measured normal case (full red line) and the case when the anomaly lasts during the total 14-day period (dashed red line). The curves are arbitrarily shifted upwards (minimum set to zero) for comparison.}
  \label{fig:compare_potential_intermittent}
\end{figure}

At last, the intermittent anomaly is considered again, yet this time the duration of the anomaly is varied, see figure \ref{fig:compare_potential_intermittent}. One sees that when the intermittent anomaly only takes place 6 hours ($1.8\%$) or less, the potential does not change. Yet if the intermittent anomaly occurs over a duration of 12 hours ($3.6\%$) or longer, the potential clearly deviates from the normal case. The deviation increases with increasing duration, as the potential approaches the full anomaly case. This type of dynamical behavior is reminiscent of phases transitions. It is similar to the cusp catastrophe described in catastrophe theory, see e.g. Ref. \cite{Zeeman1979}. Thus, the potential follows the generic law of the form $\Psi(P)=P^4+aP^2+bP$, where $a$ causes the bistability of the system. For a bistable system, the slope of the drift coefficient changes from negative to positive to negative again (instead of being only negative in the presence of only one attractive fixed point). In our case, we see how the parameter $b$ causes the phase transition by shifting $D^{(1)}$ along the $y-$axis and also causes the transition from one to two stable fixed points. This kind of bifurcation caused by the weighted mixing of two phases was also observed, e.g. for the differential resistance of some nonequilibrium semi- and super-conductor systems \cite{Peinke1987}.

\subsection{Summary}
\label{sec:monitoring_summary}
In summary, we conclude from the sensitivity analysis that the drift coefficient is a better estimate than the IEC power curve to monitor the power performance of the wind farm. It reacts faster to a change in the farm dynamics than a procedure based on ten-minute averaging. This comes from the fact that the IEC method gives an average result that deviates slowly. In contrast to this the drift estimate can detect the bistable dynamics, which provides much more insight onto the cause of the anomaly, as an emerging bistability clearly indicates that something does not function properly. We observe in scenario (a) that the drift coefficient of the entire farm changes significantly when one of the twelve turbines is shut down during $3.6\%$ of the time. Scenario (b) illustrates that the same turbine must be shut down during at least $28.6\%$ of the time for the IEC method to detect it. Additionally, scenario (c) shows that if the anomaly switches the turbine on and off intermittently, the IEC power curve can no longer detect it affirmatively. The drift coefficient detects the intermittent anomalies as well as the continuous ones, in both cases the anomaly needs only last $3.6\%$ of the time to be detected. For this reason, we promote the drift coefficient and the corresponding Langevin power curve as improved macroscopic estimates for wind farm performance monitoring.

\section{Conclusion}
\label{sec:conclu}

The conversion process of an onshore wind farm is analyzed at a sampling frequency of 1Hz. Clear dynamics appear when considering the effective wind speed $u$ and ten-minute average direction $\bar{\phi}$ averaged over all wind turbines, as well as their cumulative power production $P$. Fast power fluctuations of up to $\pm 20\%$ around the power curve are measured continuously, stressing the need for a more evolved description. Feature-rich dynamics are observed at the time scales of few seconds where the wind farm reacts to wind fluctuations. We propose a new approach where this conversion process $(u(t),\bar{\phi}) \to P(t)$ is modeled as a stochastic Langevin process. Two model coefficients are estimated from the data set measured, and describe intuitively the wind farm dynamics: the drift coefficient describes the attraction towards an attractive power curve, and the diffusion coefficient represents additional turbulent fluctuations. In addition, we introduce an additional pre-model of wind speed fluctuations, such that only ten-minute wind data is needed to model the power output at 1Hz. A statistical analysis confirms that the Langevin model reproduces the complex properties of the measurement well, including its intermittent features. Such high-frequency dynamics cannot be modeled correctly using a power curve method, that only describes the long-term dynamics. Beyond predicting the trend of the wind power production, our model predicts realistic power fluctuations at 1Hz, so that the impact of wind fluctuations on the grid stability can be forecast. Also, the added insight helps understand the dynamical response of large wind installations to wind fluctuations, such that smart grid concepts can be further refined. Thanks to their general flexible structure, such stochastic methods can be upscaled easily in order to understand and model the large wind installations that will populate future electric networks.

As a second application, we show that the drift coefficient is a compact measure of the global performance of a wind installation, making it a powerful tool to monitor the power performance of a wind farm. It captures more information about the dynamical behavior of the wind farm than the IEC power curve. For example, the drift coefficient detects a significant change in the global wind farm dynamics if one of twelve turbines is shut down during $4\%$ of the time, whereas an anomaly time of almost $30\%$ is needed for the IEC power curve to detect it. The IEC power curve is also at a disadvantage when in presence of anomalies that intermittently switch off a turbine during a short time, because the time averaging it applies strongly reduces the sensitivity to dynamical changes faster than ten minutes. On the contrary, the drift coefficient is equally reactive to intermittent and continuous anomalies. Such stochastic analysis can be useful for the real-time monitoring of wind farms. It can be applied in addition to existing tools to give some first information about the entire farm. If a potential anomaly is detected, the stochastic analysis can then be performed in greater detail for single wind turbines. If an anomaly still appears then, a full (microscopic) analysis of all the turbine sensors is justified. Also here, a stochastic analysis may be helpful too for damage detection, see Ref. \cite{Rinn2012}. Our approach is fruitful as a first tool to alert of an occurring anomaly.

\section{Acknowledgements}\label{sec:acknowledgements}
We thank the land of Lower Saxony, the German Ministry for Environment and the German Ministry for Education and Research for funding this research project. We also thank Deutsche Windtechnik AG Bremen for providing us with wind turbine data. We would also like to thank Philip Rinn, Mehrnaz Anvari, David Bastine, Benjamin Wahl, Mohammad Reza Rahimi Tabar, Allan Morales, Tanja M{\"u}cke and Michael H{\"o}lling for the many fruitful discussions.

\appendix

\section{Probabilistic description of the wind farm power production}
\label{sec:proba}

\subsection{The Fokker-Planck equation}
\label{sec:fp}

A stochastic approach was presented in section \ref{sec:model} to generate time series of power output for the wind farm. The stochastic signals are generated by solving a Langevin equation (\ref{eq:langevin}) in time. The Langevin process generated is a stochastic drift / diffusion process. The drift coefficient $D^{(1)}(P,u,\bar{\phi})$ and the diffusion coefficient $D^{(2)}(P,u,\bar{\phi})$ approximate to a first and second order how the wind farm converts a wind field $(u(t),\bar{\phi})$ into a power output $P(t)$.

The Langevin equation being by nature stochastic, the process simulated $P(t)$ has a random character. This means that if the same Langevin equation is solved $N$ times with identical wind conditions, the new signal simulated each time will always be different to some degree, owing to the random nature of the Langevin noise $\Gamma(t)$ in equation (\ref{eq:langevin}) (but its statistics won't change). While the value of $P(t)$ at a given time $t$ is random, the probability $f(P,t)$ of having a power value $P$ at time $t$ is fixed. The probability density function of the Langevin process $P(t)$ at any given time $t$ is uniquely defined as the solution of the Fokker-Planck equation \cite{Risken1996,Friedrich2011}
\begin{align}
\frac{\partial f(P(t),t)}{\partial t}=-\frac{\partial}{\partial P(t)} &\Big( D^{(1)}\big(P(t),u(t),\bar{\phi}\big) \cdot f\big(P(t),t\big) \Big) \nonumber \\ &+ \frac{1}{2}\frac{\partial^2}{\partial P(t)^2} \Big(D^{(2)}\big(P(t),u(t),\bar{\phi}\big) \cdot f\big(P(t),t\big) \Big) \, ,
\label{eq:fp}
\end{align}
where the Kramers-Moyal coefficients are estimated following equation (\ref{eq:kmc}). The Fokker-Planck equation is the equivalent in the probability domain of the Langevin equation in the time domain.

\begin{figure}[!h]
   \centering
   \begin{minipage}[t]{0.48\linewidth}
      \centering
   	 \includegraphics[width=0.8\columnwidth]{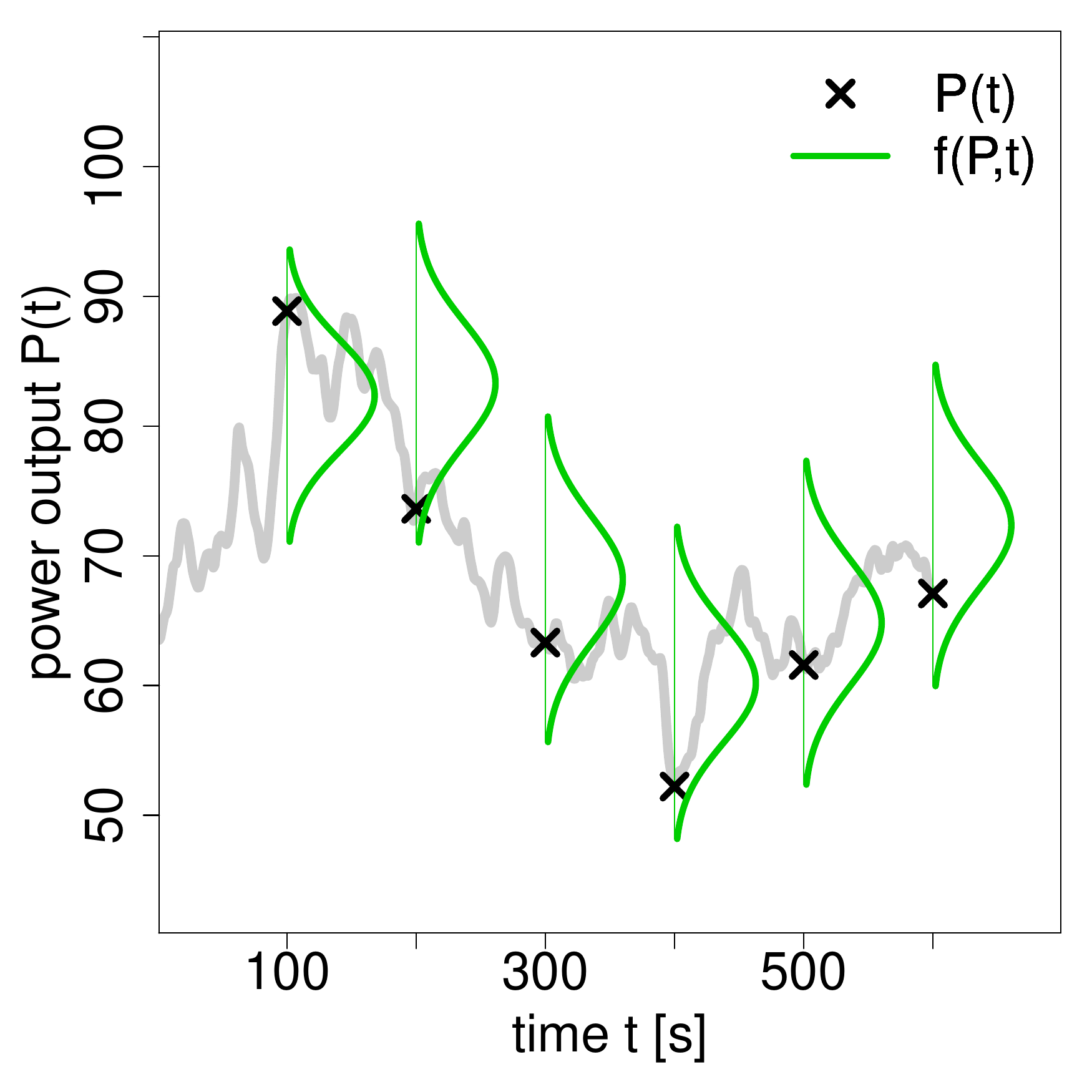}  
 	\caption{PDF $f(P,t)$ (green curve) at times $t=[100,200,300,400,500,600]$s calculated from equation (\ref{eq:fp}). The power value measured at those times is indicated (black cross). The power output measured $P(t)$ is also shown (gray line).}
   \label{fig:L_FP}
    \end{minipage}%
   \hspace{0.5cm}%
   \begin{minipage}[t]{0.48\linewidth}
      \centering
   	 \includegraphics[width=0.8\columnwidth]{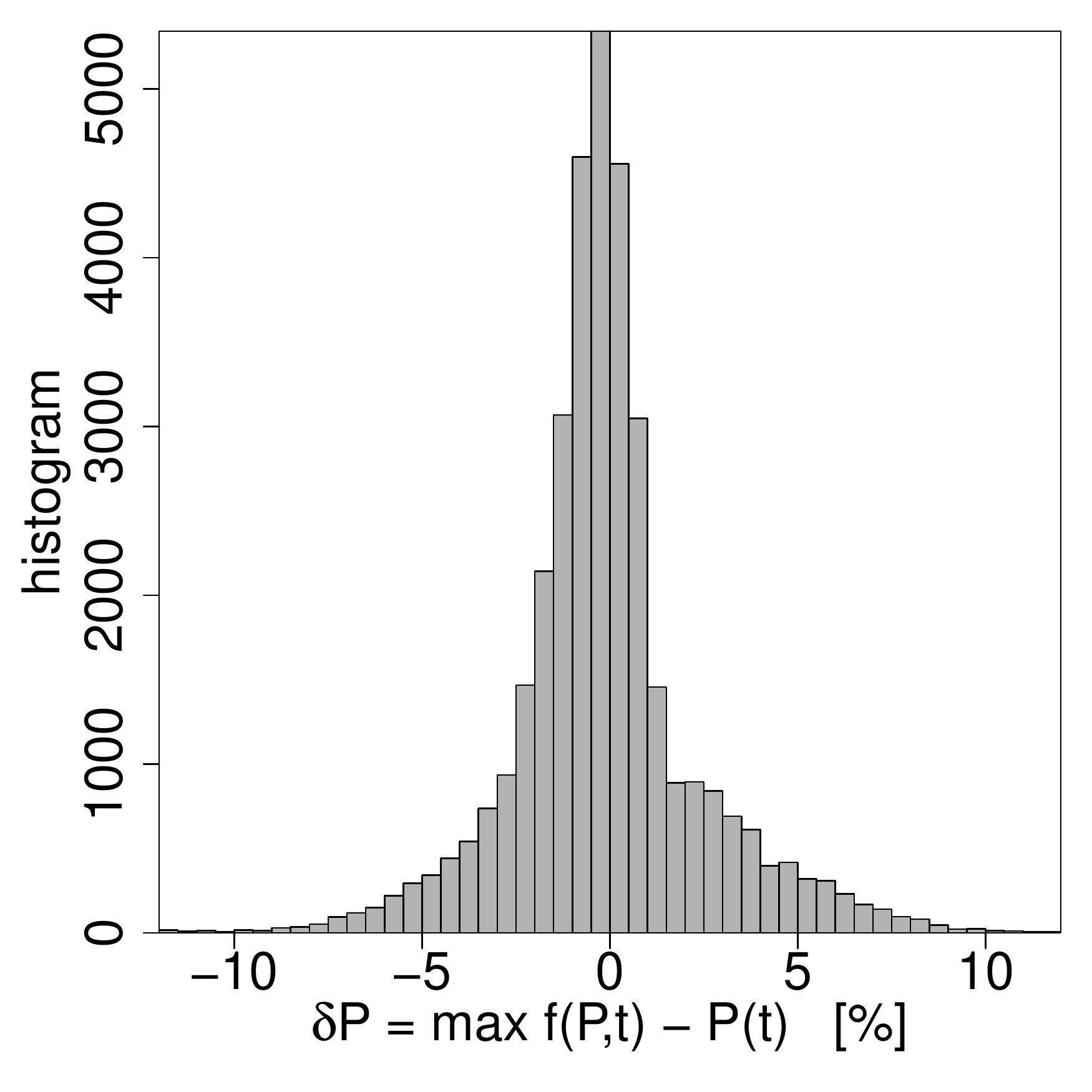}  
 	\caption{Histogram of the difference $\delta P = max \, f(P,t) - P(t)$ estimated over 36000 samples (10h).}
   \label{fig:FP_deviation}
   \end{minipage}
\end{figure}

If the probability $f(P,t)$ is desired instead of simply one sample value $P(t)$, the Fokker-Planck equation should be solved over time \footnote{We solve the Fokker-Planck equation using the {\it path integral method} presented in Ref. \cite{Risken1996}. Given an initial condition $f(P,t=0)$, we calculate $f(P,t>0)$ by repeatedly solving the equation $f(P,t+dt)=\int f(P,t+dt|P',t) f(P',t) dP'$ over an infinitesimal time step $dt$, where the conditional probability $f(P,t+dt|P',t)$ is completely described by $D^{(1)}$ and $D^{(2)}$. One can visualize how the drift and diffusion coefficients concurrently contract and stretch the PDF over time.}. An example is presented in figure \ref{fig:L_FP} where the Fokker-Planck equation is solved at 1Hz using as initial condition a Dirac distribution $f(P,t=0)=\delta(P(t=0))$. We see at six exemplary times $t$ what the measured power value $P(t)$ is, and the probability $f(P,t)$ estimated by the Fokker-Planck equation. 

The difference $\delta P$ between the most probable power value given by the Fokker-Planck equation $max\,f(P,t)$ and the measurement $P(t)$ is calculated for 36000 samples, and its histogram is presented in figure \ref{fig:FP_deviation}. $\delta P$ is rather centered around zero, has a mean value $-0.11\%$ and a standard deviation $2.54\%$. The standard deviation of $P(t)$ is equal to $\sigma_P=22.63\%$, that is 9 times as large as the standard deviation of the difference. This means that the most probable power value predicted by the Fokker-Planck equation $max\,f(P,t)$ deviates from the measurement by only $\sigma_P/9$ on average. This concludes that the PDF simulated by the Fokker-Planck equation follows the measurement faithfully.

\subsection{Probability-based optimization of the Kramers-Moyal coefficients}
\label{sec:opt}

\begin{figure} 
  	\begin{center}
   	 \includegraphics[width=0.5\columnwidth]{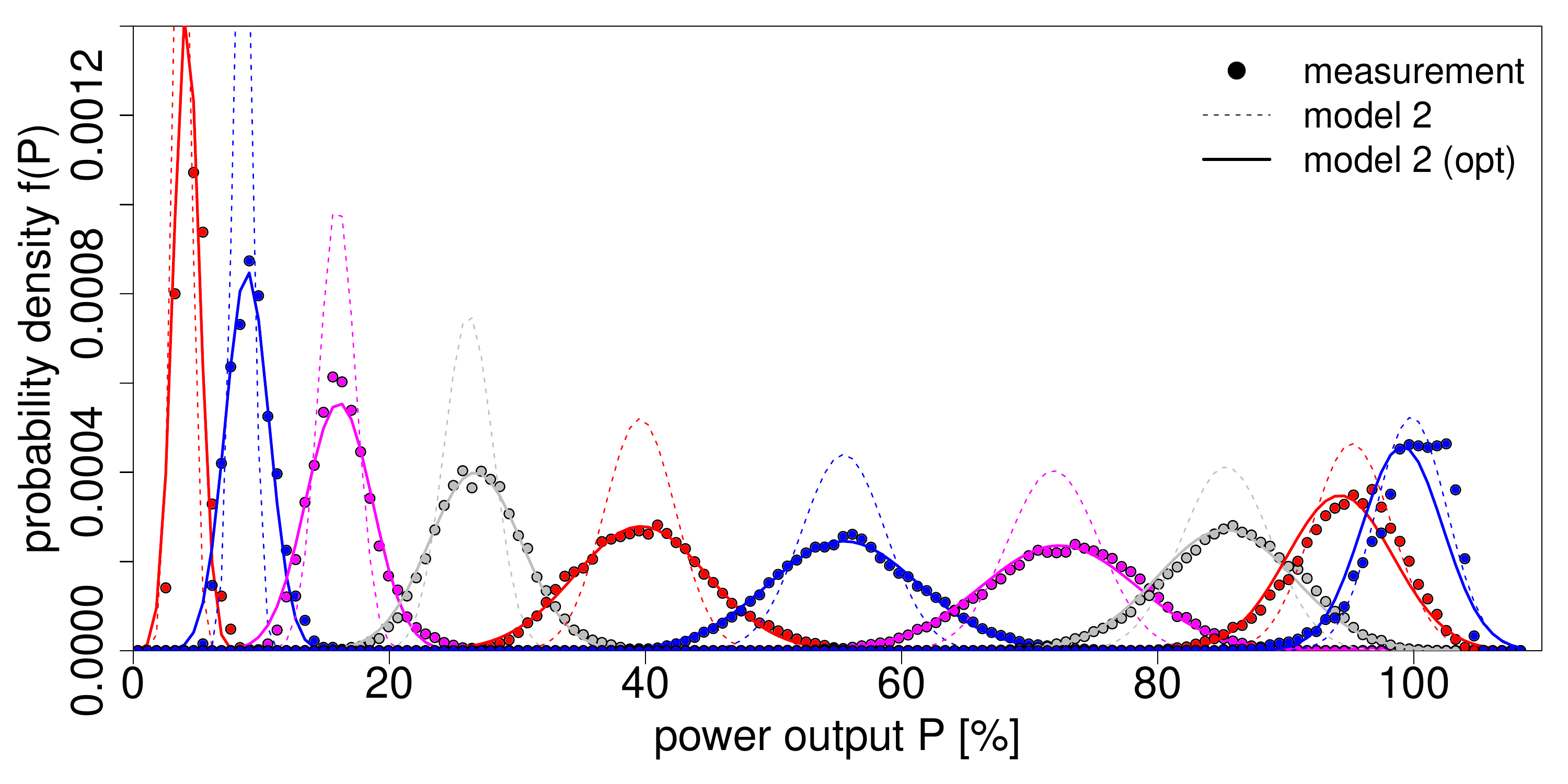}  
	\end{center}
	\caption{PDF of measured power $f(P|u_j,\bar{\phi}_k)$ (dots) and PDF $f_{st}(P)$ generated by model 2 (dashed line) and model 2 optimized (bold full line) for various wind speed bins $j \in [9,27]$ (various colors) for the wind sector $k=8$ ($\bar{\phi}\simeq 225^\circ$).}
  \label{fig:power_histo_model2}
\end{figure}

The drift coefficient $D^{(1)}(P,u,\bar{\phi})$ and the diffusion coefficient $D^{(2)}(P,u,\bar{\phi})$ are defined in each bin of the three-dimensional phase space $\{P_i,u_j,\bar{\phi}_k\}$. A parametric form is presented in subsection \ref{sec:model2} where in each sub-domain $\{u_j,\bar{\phi}_k\}$, the drift is simplified to a linear function $D^{(1)}(P,u_j,\bar{\phi}_k) \simeq \alpha_{jk} \cdot ( P - P_{L}(u_j,\bar{\phi}_k) )$ and the diffusion to a constant $D^{(2)}(P,u_j,\bar{\phi}_k) \simeq \beta_{jk}$. The two parameters $\alpha_{jk}$ and $\beta_{jk}$ are fitted from the drift and diffusion coefficients estimated. However, some artifacts such as low data sampling or the presence of measurement noise within the data affect the estimated Kramers-Moyal coefficients, which then deviate from the true coefficients of the system. We propose an optimization procedure that corrects this discrepancy by refining the estimation of $D^{(1)}$ and $D^{(2)}$.

As long as the system stays inside the given sub-domain $\{u_j,\bar{\phi}_k\}$, the local dynamics are dictated by the local values of drift and diffusion. During that period of time, the dynamics do not change and the system fluctuates around the power curve value $P_{L}(u_j,\bar{\phi}_k)$. This corresponds to the PDF $f(P,t)$ relaxing to a stationary form $f_{st}(P)$ that is given by
\begin{eqnarray}
0 &=& \frac{\partial f_{st}(P)}{\partial t} \\
0 &=& -\Bigg( \frac{\partial}{\partial P(t)} D^{(1)}\big(P(t),u(t),\bar{\phi}\big) \cdot f_{st}\big(P\big) \Bigg) + \frac{1}{2}\Bigg(\frac{\partial^2}{\partial P(t)^2} D^{(2)}\big(P(t),u(t),\bar{\phi}\big) \cdot f_{st}\big(P\big) \Bigg) \, ,
\label{eq:fp_zero}
\end{eqnarray}
giving
\begin{eqnarray}
f_{st}(P)=\frac{N}{D^{(2)}\big(P,u,\bar{\phi}\big)} \,\, exp \Bigg( \int^P \frac{D^{(1)}\big(P',u,\bar{\phi}\big)}{D^{(2)}\big(P',u,\bar{\phi}\big)} dP' \Bigg) \, ,
\label{eq:fp_stat}
\end{eqnarray}
where $N$ is a normalization constant such that $\int_{-\infty}^{\infty} f_{st}(P) \, dP=1$. $f_{st}(P)$ is the stationary solution of the Fokker-Planck equation, that corresponds to the stationary case where $D^{(1)}$ and $D^{(2)}$ do not change in time. In the limit of a system that reacts infinitely fast to the changing wind condition, one can consider that between two changes of the wind condition, the system has time to relax to its stationary state and wait for the next change. We assume that our system is reasonably close to this limit, owing to the low dispersion of the data around the power curve (see figure \ref{fig:uP_1Hz}). This means that all the data $(P|u_j,\bar{\phi}_k)$ contained in the sub-domain $\{u_j,\bar{\phi}_k\}$ is assumed to have relaxed towards its local power curve value $P_{L}(u_j,\bar{\phi}_k)$, and its PDF is
\begin{eqnarray}
f(P|u_j,\bar{\phi}_k) \sim f_{st}(P)=\frac{N}{\beta_{jk}} \,\, exp \Bigg( - \frac{\alpha_{jk} \big(P-P_{L}(u_j,\bar{\phi}_k)\big)^2}{2 \, \beta_{jk}} \Bigg) \, .
\label{eq:fp_stat_opt}
\end{eqnarray} 
The two PDFs $f(P|u_j,\bar{\phi}_k)$ and $f_{st}(P)$ are presented in figure \ref{fig:power_histo_model2} for various wind speed bins. We observe a reasonably good agreement between the two PDFs, which can be improved. The optimization is based on the fact that the PDFs are reasonably close to a Gaussian distribution
\begin{eqnarray}
f(P|u_j,\bar{\phi}_k) \simeq \frac{1}{\sqrt{2\pi \sigma_P \, ^2}} \,\, exp \Bigg( - \frac{ \big(P-\bar{P})\big)^2}{2 \, \sigma_P \, ^2} \Bigg) \, ,
\label{eq:fp_stat_opt_gauss}
\end{eqnarray}
with $\bar{P}$ and $\sigma_P$ the mean and standard deviation of $(P|u_j,\bar{\phi}_k)$. Observing the analogy between equations (\ref{eq:fp_stat_opt}) and (\ref{eq:fp_stat_opt_gauss}), we optimize the drift and diffusion coefficients following
\begin{eqnarray}
D^{(1)}_{opt}(P,u_j,\bar{\phi}_k) &=& \alpha_{jk} \cdot ( P - \bar{P} )\\
D^{(2)}_{opt}(P,u_j,\bar{\phi}_k) &=& \alpha_{jk} \cdot \sigma_P \,^2 \, .
\label{eq:kmc_opt}
\end{eqnarray}
The stationary PDF calculated from the optimized coefficients is presented in figure \ref{fig:power_histo_model2}. We observe a better agreement to the measured PDFs, validating the optimization procedure \footnote{A chi-square test was performed to quantify the deviation between measured and reconstructed PDFs. When averaging the chi-square value $\chi^2_j$ over all wind speed bins $j$, we find for the average $< \chi^2_j >_j$ that the optimization reduces the value by a factor $6.8$.}. To summarize, we relate the stationary solution of the Fokker-Planck equation to the PDF of the measured power data, in each wind speed/direction sub-domain. This way we can fine-tune the parametric form of the drift and diffusion coefficients to reproduce exactly the distribution of the measured data. This optimization is used for models 2 and 3 presented resp. in subsections \ref{sec:model2} and \ref{sec:model3}.


\begin{thebibliography}{10}

\bibitem{WWEA2010}
World Wind~Energy Association.
\newblock World wind energy report 2010.
\newblock Technical report, World Wind Energy Association, 2011.

\bibitem{Liu2011}
J.~Liu, B.~H. Krogh, and B.~E. Ydstie.
\newblock Passivity-based robust control for power systems subject to wind
  power variability.
\newblock In {\em American Control Conference}, 2011.

\bibitem{Ayodele2012}
TR~Ayodele, AA~Jimoh, JL~Munda, and JT~Agee.
\newblock The impact of wind power on power system transient stability based on
  probabilistic weighting method.
\newblock {\em Journal of Renewable and Sustainable Energy}, 4, 2012.

\bibitem{Milan2013a}
Patrick Milan, Matthias W\"achter, and Joachim Peinke.
\newblock Turbulent character of wind energy.
\newblock {\em Phys. Rev. Lett.}, 110:138701, Mar 2013.

\bibitem{Rohden2012}
Martin Rohden, Andreas Sorge, Marc Timme, and Dirk Witthaut.
\newblock Self-organized synchronization in decentralized power grids.
\newblock {\em Phys. Rev. Lett.}, 109:064101, Aug 2012.

\bibitem{Liu2013}
Dewei Liu, Jianbo Guo, Yuehui Huang, and Weisheng Wang.
\newblock An active power control strategy for wind farm based on predictions
  of wind turbine's maximum generation capacity.
\newblock {\em Journal of Renewable and Sustainable Energy}, 5:013121, 2013.

\bibitem{IEC}
{IEC 61400-1}.
\newblock {\em Wind turbines. Part 1: Design requirements}.
\newblock International Electrotechnical Commission, 2005.
\newblock {IEC}/61400-1, 3rd Edition.

\bibitem{Lovejoy2001}
S.~Lovejoy, D.~Schertzer, and J.~D. Stanway.
\newblock Direct evidence of multifractal atmospheric cascades from planetary
  scales down to 1 km.
\newblock {\em Phys. Rev. Lett.}, 86(22):5200--5203, May 2001.

\bibitem{Boettcher2007a}
F.~B{\"o}ttcher, J.~Peinke, D.~Kleinhans, and R.~Friedrich.
\newblock Handling systems driven by different noise sources -- implications
  for power estimations.
\newblock In {\em Wind Energy}, pages 179--182. Springer, Berlin, 2007.

\bibitem{Morales2010a}
A~Morales, M~W{\"a}chter, and J~Peinke.
\newblock Characterization of wind turbulence by higher-order statistics.
\newblock {\em Wind Energy}, 15(3):391--406, 2012.

\bibitem{Lovejoy2009}
S.~Lovejoy, D.~Schertzer, V.~Allaire, T.~Bourgeois, S.~King, J.~Pinel, and
  J.~Stolle.
\newblock Atmospheric complexity or scale by scale simplicity.
\newblock {\em GRL}, 36:L01801, January 2009.

\bibitem{Bianchi2006}
F.D. Bianchi, H.~{De Battista}, and R.J. Mantz.
\newblock {\em Wind Turbine Control Systems}.
\newblock Springer, Berlin, 2nd edition, 2006.

\bibitem{Tavner2011}
Peter Tavner, Yingning Qiu, Athanasios Korogiannos, and Yanhui Feng.
\newblock The correlation between wind turbine turbulence and pitch failure.
\newblock In {\em Proceedings of EWEA 2011}, 2011.

\bibitem{Muecke2011}
Tanja M{\"u}cke, David Kleinhans, and Joachim Peinke.
\newblock Atmospheric turbulence and its influence on the alternating loads on
  wind turbines.
\newblock {\em Wind Energy}, 14(2):301--316, 2011.

\bibitem{vdHoven1957}
Isaac Van~der Hoven.
\newblock Power spectrum of horizontal wind speed in the frequency range from
  0.0007 to 900 cylcles per hour.
\newblock {\em Journal of Meteorology}, 14(2):160--164, 1957.

\bibitem{Schertzer2012}
D.~Schertzer, I.~Tchiguirinskaia, S.~Lovejoy, and A.~F. Tuck.
\newblock Quasi-geostrophic turbulence and generalized scale invariance, a
  theoretical reply.
\newblock {\em Atmospheric Chemistry and Physics}, 12(1), 2012.

\bibitem{Fitton2011}
G.~Fitton, I.~Tchiguirinskaia, D.~Schertzer, and S.~Lovejoy.
\newblock The anisotropic multifractal model and wind turbine wakes.
\newblock In {\em 7th PhD Seminar on Wind Energy in Europe}, 2011.

\bibitem{Kaminsky1991}
F.~C. Kaminsky, R.~H. Kirchhoff, C.~Y. Syu, and J.~F. Manwell.
\newblock A comparison of alternative approaches for the synthetic generation
  of a wind speed time series.
\newblock {\em Journal of Solar Energy Engineering}, 113(4):280--289, 1991.

\bibitem{Muzy2010b}
Rachel Ba{\"i}le, Philippe Poggi, and Jean-Francois Muzy.
\newblock Intermittency model for surface layer wind speed fluctuations.
  applications to short term forecasting and calibration of the wind resource.
\newblock In {\em Proceedings of EWEC}, 2010.

\bibitem{Calif2012a}
Rudy Calif.
\newblock Pdf models and synthetic model for the wind speed fluctuations based
  on the resolution of langevin equation.
\newblock {\em Applied Energy}, 99(0):173 -- 182, 2012.

\bibitem{Calif2012b}
R.~Calif and F.G. Schmitt.
\newblock Modeling of atmospheric wind speed sequence using a lognormal
  continuous stochastic equation.
\newblock {\em Journal of Wind Engineering and Industrial Aerodynamics},
  109(0):1 -- 8, 2012.

\bibitem{IEC-12}
{IEC 61400-12}.
\newblock {\em Wind turbines - Part 12: Power performance measurements of
  electricity producing wind turbines}.
\newblock International Electrotechnical Commission, 2005.

\bibitem{Burton2001}
T.~Burton, D.~Sharpe, N.~Jenkins, and E.~Bossanyi.
\newblock {\em Wind energy handbook}.
\newblock Wiley, New York, 2001.

\bibitem{Wu2014}
Bingheng Wu, Mengxuan Song, Kai Chen, Zhongyang He, and Xing Zhang.
\newblock Wind power prediction system for wind farm based on auto regressive
  statistical model and physical model.
\newblock {\em Journal of Renewable and Sustainable Energy}, 6(1):013101, 2014.

\bibitem{Kamps2012}
Oliver Kamps.
\newblock Characterizing the fluctuations of wind power production by
  multi-time statistics.
\newblock In {\em Proceedings of the Euromech Colloquium 528}, 2012.

\bibitem{Muzy2010a}
Jean-Francois Muzy, Rachel Ba{\"i}le, and Philippe Poggi.
\newblock Intermittency of surface-layer wind velocity series in the mesoscale
  range.
\newblock {\em Phys. Rev. E}, 81(5):056308, May 2010.

\bibitem{Blanco2009}
Mar{\'\i}a~Isabel Blanco.
\newblock The economics of wind energy.
\newblock {\em Renewable and Sustainable Energy Reviews}, 13(6):1372--1382,
  2009.

\bibitem{Hahn2007}
Berthold Hahn, Michael Durstewitz, and Kurt Rohrig.
\newblock Reliability of wind turbines.
\newblock In {\em Wind energy}, pages 329--332. Springer, 2007.

\bibitem{IWES2012}
Sebastian Pfaffel, Volker Berkhout, Stefan Faulstich, Paul K{\"u}hn, Katrin
  Linke, Philipp Lyding, and Renate Rothkegel.
\newblock Wind energy report germany 2011.
\newblock Technical report, Fraunhofer Institute for Wind Energy and Energy
  System Technology (IWES), 2012.

\bibitem{Ciang2008}
Chia~Chen Ciang, Jung-Ryul Lee, and Hyung-Joon Bang.
\newblock Structural health monitoring for a wind turbine system: a review of
  damage detection methods.
\newblock {\em Measurement Science and Technology}, 19(12):122001, 2008.

\bibitem{Cutler2011}
Nicholas~J. Cutler, Hugh~R. Outhred, and Iain~F. MacGill.
\newblock Using nacelle-based wind speed observations to improve power curve
  modeling for wind power forecasting.
\newblock {\em Wind Energy}, pages 245--258, 2011.

\bibitem{Milan2011a}
Patrick Milan, Matthias W{\"a}chter, and Joachim Peinke.
\newblock Stochastic modeling of wind power production.
\newblock In {\em Proceedings of EWEA 2011}, Brussels, 2011.

\bibitem{Anahua2007}
Edgar Anahua, Stephan Barth, and Joachim Peinke.
\newblock Markovian power curves for wind turbines.
\newblock {\em Wind Energy}, 11(3), 2008.

\bibitem{Gottschall2008}
Julia Gottschall and Joachim Peinke.
\newblock How to improve the estimation of power curves for wind turbines.
\newblock {\em Environmental Research Letters}, 3(1):015005 (7pp), 2008.

\bibitem{Risken1996}
Hannes Risken.
\newblock {\em The {F}okker-{P}lanck Equation}.
\newblock Springer, 1996.

\bibitem{Friedrich2011}
Rudolf Friedrich, Joachim Peinke, Muhammad Sahimi, and M.~Reza~Rahimi Tabar.
\newblock Approaching complexity by stochastic methods: From biological systems
  to turbulence.
\newblock {\em Physics Reports}, 2011.

\bibitem{Boettcher2006}
Frank B{\"o}ttcher, Joachim Peinke, Rudolf Friedrich, David Kleinhans, Pedro~G.
  Lind, and Maria Haase.
\newblock Reconstruction of complex dynamical systems affected by strong
  measurement noise.
\newblock {\em Phys. Rev. Lett.}, 97, 2006.

\bibitem{Sura2002}
Philip Sura and Joseph Barsugli.
\newblock A note on estimating drift and diffusion parameters from timeseries.
\newblock {\em Physics Letters A}, 305:304--311, 2002.

\bibitem{Gottschall2008b}
J.~Gottschall and J.~Peinke.
\newblock On the definition and handling of different drift and diffusion
  estimates.
\newblock {\em New J. Phys.}, 10:083034 (20pp), 2008.

\bibitem{Haken1983}
H~Haken.
\newblock Advanced synergetic, instability hierarchies of self-organizing
  systems and devices, 1983.

\bibitem{Boettcher2007}
Frank B{\"o}ttcher, Stephan Barth, and Joachim Peinke.
\newblock Small and large fluctuations in atmospheric wind speeds.
\newblock {\em Stoch Environ Res Ris Assess}, 21:299--308, 2007.

\bibitem{Stresing2010}
Robert Stresing and J~Peinke.
\newblock Towards a stochastic multi-point description of turbulence.
\newblock {\em New Journal of Physics}, 12(10):103046, 2010.

\bibitem{Press2007}
William~H. Press, Saul~A. Teukolsky, William~T. Vetterling, and Brian~P.
  Flannery.
\newblock {\em Numerical Recipes: The Art of Scientific Computing}.
\newblock Cambridge University Press, third edition, 2007.

\bibitem{Arneodo1996}
A.~Arneodo, C.~Baudet, F.~Belin, R.~Benzi, B.~Castaing, B.~Chabaud,
  R.~Chavarria, S.~Ciliberto, R.~Camussi, F.~Chilla, B.~Dubrulle, Y.~Gagne,
  B.~Hebral, J.~Herweijer, M.~Marchand, J.~Maurer, J.~F. Muzy, A.~Naert,
  A.~Noullez, J.~Peinke, F.~Roux, P.~Tabeling, W.~van~de Water, and
  H.~Willaime.
\newblock Structure functions in turbulence, in various flow configurations, at
  reynolds number between 30 and 5000, using extended self-similarity.
\newblock {\em EPL (Europhysics Letters)}, 34(6), 1996.

\bibitem{Benzi1993}
R.~Benzi, S.~Ciliberto, R.~Tripiccione, C.~Baudet, F.~Massaioli, and S.~Succi.
\newblock Extended self-similarity in turbulent flows.
\newblock {\em Physical Review E}, 48(1), 1993.

\bibitem{Kavasseri2004}
R.G. Kavasseri and R.~Nagarajan.
\newblock Evidence of crossover phenomena in wind speed data.
\newblock {\em IEEE Transactions on Circuits and Systems}, 51 (11):2255--2262,
  November 2004.

\bibitem{Zeeman1979}
E~Christopher Zeeman.
\newblock {\em Catastrophe theory}.
\newblock Springer, 1979.

\bibitem{Peinke1987}
J~Peinke, DB~Schmid, B~R{\"o}hricht, and J~Parisi.
\newblock Positive and negative differential resistance in electrical
  conductors.
\newblock {\em Zeitschrift f{\"u}r Physik B Condensed Matter}, 66(1):65--73,
  1987.

\bibitem{Rinn2012}
P.~Rinn, H.~Hei{\ss}elmann, M.~W{\"a}chter, and J.~Peinke.
\newblock Stochastic method for in-situ damage analysis.
\newblock {\em The European Physical Journal B}, 86(1):1--5, 2012.

\end{thebibliography}
\end{document}